\documentclass[fleqn,usenatbib]{mnras}
\usepackage{newtxtext,newtxmath}

\usepackage[T1]{fontenc}
\usepackage{ae,aecompl}

\usepackage{blindtext}
\usepackage{siunitx}
\DeclareSIUnit\mpc{Mpc}

\usepackage{graphicx}	
\usepackage{amsmath}	

\title[Quenched and transient galaxies]{The environmental dependence of rapidly-quenching and rejuvenating galaxies} 

\author[C. Cleland \& S. L. McGee]{
Cressida Cleland,\thanks{E-mail: cressidac@star.sr.bham.ac.uk}
Sean L. McGee
\\
School of Physics and Astronomy, University of Birmingham, Edgbaston, Birmingham, B15 2TT, UK\\
}

\date{Accepted XXX. Received YYY; in original form ZZZ}

\pubyear{2019}

\begin{document}
\label{firstpage}
\pagerange{\pageref{firstpage}--\pageref{lastpage}}
\maketitle

\begin{abstract}
By combining H$\upalpha$ flux measurements from the Sloan Digital Sky Survey (SDSS) with UV flux observations from the Galaxy Evolution Explorer (GALEX), we examine the environmental dependence (through central/satellite distinction) of the rapid quenching and rejuvenation of galaxies. H$\upalpha$ emissions trace the most massive stars, thereby indicating star-formation on timescales of $\sim$ 10 Myr, while UV emission traces star-formation on timescales of $\sim$ 100 Myr. 
These varying timescales are exploited to probe the most recent star-formation histories of galaxies. In this work, we define a class of transient galaxies which have UV emission typical of star-formation but negligible H$\upalpha$ emission. We find that the occurrence of these transients has a strong stellar mass dependence in both the satellite and central population. However, while at stellar masses greater than $M_*\sim 10^{10}$ M$_\odot$ they occur with equal frequency regardless of environmental class, at lower stellar masses they are more common in satellites only, with an excess of about 1 percentage point across all low stellar mass galaxies. These satellite transients also have a strong halo mass and group-centric radial dependence suggesting they are driven by an environmental process. Finally, we select a sample of galaxies with H$\upalpha$ emission but not UV emission which could contain short-timescale rejuvenating galaxies. These rejuvenating candidates are few in number and do not have a strong difference in their occurrence rate in centrals or satellites. These unique probes point to an environmental quenching mechanism which occurs on short timescales after the satellite has been in the group environment for a significant time -- consistent with `delayed-then-rapid' quenching.
\end{abstract}




\section{Introduction}

It has been established that the distribution of colours of galaxies is bimodal, with star-forming galaxies belonging to the `blue cloud' and quiescent galaxies belonging to the `red sequence' \citep{strateva2001,bell2004}. The redshift evolution of galaxy populations suggests that the quenching of star-formation is the likely process which moves a galaxy from the blue cloud to the red sequence \citep{baldry2004,faber2007}. However, there are many separate physical mechanisms that may be contributing to the reduction and cessation of star-formation in a galaxy. These mechanisms are often broadly classed as internal or external processes. \par

 \begin{figure*}
     \centering
     \includegraphics[width=\textwidth]{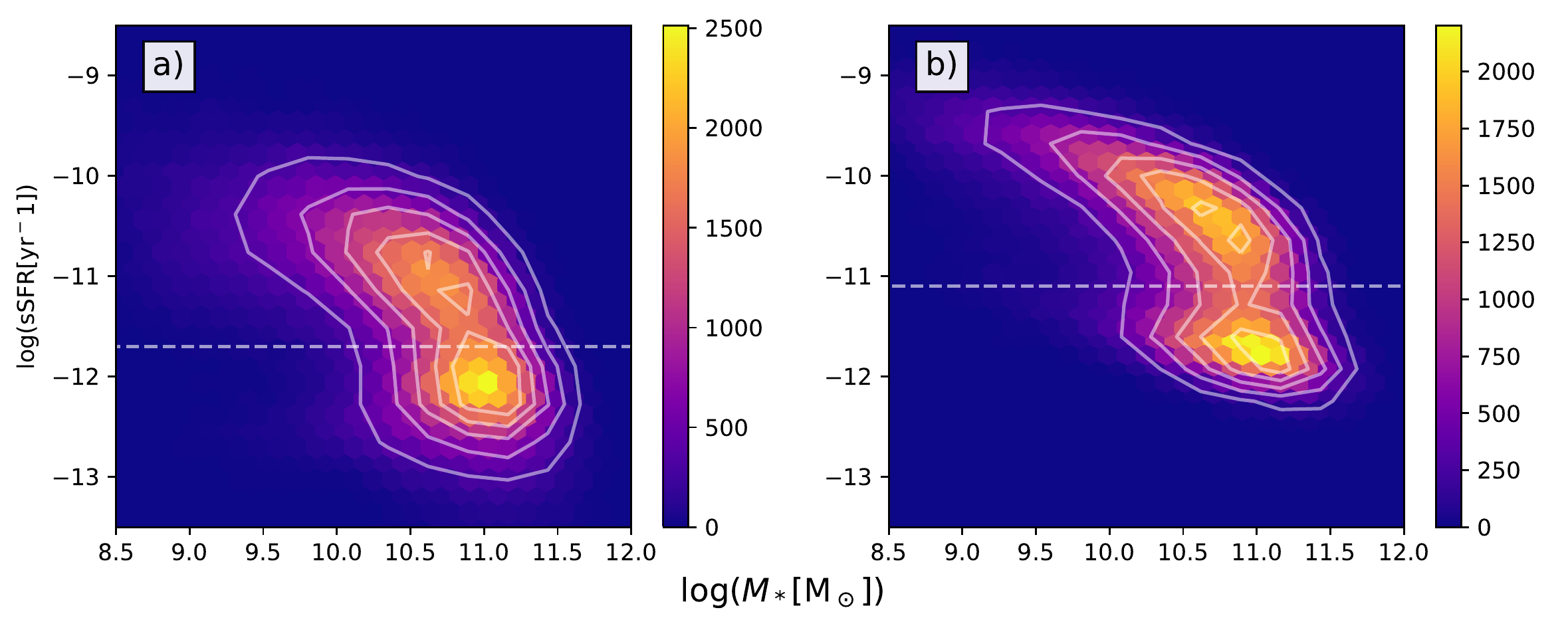}
     \caption{Stellar mass ($M*$) versus specific star-formation rate [a) $\text{sSFR}_{\text{H}\upalpha}$; b) $\text{sSFR}_{\text{UV}}$] distributions for all galaxies. Note the bimodality of the population: star-forming [$\log(\text{sSFR}_{\text{H}\upalpha}/\text{yr}^{-1})>-11.7$; $\log(\text{sSFR}_{\text{UV}}/\text{yr}^{-1})>-11.1$] and quenched [$\log(\text{sSFR}/\text{yr}^{-1})<-11.7$; $\log(\text{sSFR}_{\text{UV}}/\text{yr}^{-1})<-11.1$]. The $z$-axis denotes number of galaxies in each bin, with contour lines indicating regions of equal numbers. The dark blue regions signify no data.}
     \label{fig:uv_ha_hexbin}
 \end{figure*}
Internal processes refer to those that occur due to the individual galaxy properties, such as morphology or the presence of an AGN. Morphological quenching may either prevent star-formation from taking place, such as with the presence of a bulge which may prevent the collapse of the disc and the formation of stars \citep{bluck14}, or accelerate gas consumption, whereby the presence of a bar directs material to the centre of the galaxy \citep{athanassoula92a} where it is depleted more efficiently thus leading to a reduced SFR \citep{zurita2004,sheth2005}. Alternatively, a galaxy with an AGN may experience feedback which provides heat to the surrounding gas, preventing star-formation \citep{smethurst2017}.\par
 
External processes are those where the environment of a galaxy is relevant. These include mergers, which may violently heat and expel gas, and are also often responsible for morphological changes to a galaxy \citep{gabor2010}. Ram-pressure stripping occurs when a galaxy encounters gas upon infall to a group or cluster, causing its cool gas to be stripped away. The pressure on the galaxy is as $P_\text{R} \approx \rho_e v^2$, where $\rho_e$ is the external intracluser density and $v$ is the velocity of the galaxy \citep{gunngott1972}. Ram-pressure stripping results in a relatively quick depletion of gas, operating on crossing-time timescales \citep[$\sim 10^7$ yrs,][]{abadi1999}. A galaxy will quench its star-formation on gas consumption timescales if it fails to accrete additional cool gas through either the removal of the gas reservoir \citep[e.g. `strangulation',][]{balogh2000} or the cutting off of cosmological accretion \citep[e.g. `overconsumption',][]{mcgee2014}.\par

It is not trivial to discern which of these internal or external processes are dominant (if any) for a given galaxy as they are heavily related and dependent on each other. One way to determine if internal or external processes are dominant is by examining the galaxy properties as a function of their environment. If a particular property has no environmental dependence, then it is likely driven by internal mechanisms, and alternatively, a strong environmental dependence is indicative of a strong external mechanism. There is clear evidence that local environment of a galaxy has a prominent effect on its star-formation rate. In addition to a well-documented morphology-environment correlation \citep{dressler80}, it has been shown that satellite galaxies in a group or cluster experience an increased rate of quenching compared to galaxies at the centre of the group or cluster or isolated field galaxies \citep[e.g.,][]{mcgee2011,wetzel2012,ziparo2014}. However, clear evidence for the regimes in which internal or external mechanisms dominate is not yet clear. One promising avenue for further elucidation of this is to examine the locations and abundances of galaxies currently undergoing rapid quenching. \par

Much of the work on rapidly quenching galaxies has been done using the class of galaxies called E+A or post-starburst galaxies. These were first discovered by \citet{dressler1983} who found three galaxies within a galaxy cluster with very strong Balmer absorption lines. Using stellar population modelling, those authors suggested these galaxies resulted from a burst of star-formation ($\sim 20$ per cent of the total stellar mass formed within 10$^8$ yrs), and later became quenched. This early identification and interpretation has led to their synonymous identification as `post-starburst' galaxies. Later work focused on the occurrence of these galaxies in clusters, and their classification standardized into requiring the non-existence of a current star-formation indicator ([OII] in early studies) along with the strong H$\updelta$ absorption. It has been clear since the early days, when the proportion of cluster galaxies classified as E+A's varied from $\sim 30$ per cent to $2$ per cent \citep{couch1987, poggianti1999, balogh1999},  that the relatively small sample sizes of E+A's combined with their sensitive dependence on the details of the target selection, spectroscopic reduction and parameter estimation, could have a big impact on the interpretation. These uncertainties have continued, and the literature is unsettled on many of the major questions about E+A environmental dependencies \citep{zabludoff1996, paccagnella2019, owers2019}, morphological composition \citep{pracy2009, swinbank2012, french2016} and triggering mechanisms \citep{zabludoff1996, bekki2014, yesuf2014, pawlik2019, vulcani2020}.

In this study, we begin a fresh look at the question of rapidly quenching galaxies by selecting them in a new way -- using a combination of UV emission and H$\upalpha$ emission. Although some studies have examined the UV properties of previously selected E+A galaxies \citep{kaviraj2007, melnick2013}, to our knowledge, this is the first time where the primary selection of the rapidly quenching galaxies has been made using UV with spectral diagnostics. Our method has the benefit of being sensitive to shorter timescales of star-formation than the traditional E+A method, and thus may allow a better pinpointing of the quenching driver. We will constrain the timescale of the quenching processes by determining the recent star-formation history (SFH) of each galaxy. To do this, we define a `transient' galaxy as one that has a star-forming UV component but a lack of H$\upalpha$. H$\upalpha$ traces young O-type stars, which typically have a lifetime of about 10 Myr, while UV traces O- and B-type stars, where the latter typically lives for $\sim$100 Myr. This disparity means that a lack of H$\upalpha$ measurements implies star-formation has ceased less than $\sim$100 Myr ago. Due to the sensitivity of these star-formation tracers to stars of different masses, this method allows us to investigate the short timescale differential SFR of galaxies in the process of quenching.

Using this new classification of galaxy, we will attempt to isolate the internal/external mechanisms by examining the differential effects of quenching on the environment of the galaxy and by characterizing the timescale over which the cessation of star-formation occurs. We define `environment' as the location of the galaxy in its host dark matter halo. Central galaxies are the galaxies that have the most stellar mass in their local potential and are often assumed to be located in the center of the dark matter halo. By this definition, central galaxies contain both the dominant galaxy in a group or cluster as well as an isolated galaxy in the `field'. Galaxies which appear gravitationally bound to a halo in which they are not the dominant galaxy are called `satellites'. \par

Subsequently, the properties of each satellite and how it relates to its group may be investigated to probe the effects of the immediate environment of the satellite. For this work, we consider the group-centric radial position of satellite galaxies, and the relative velocity of the satellite to the central galaxy. Combining all these data, we aim to discern the dependencies of these quenching processes and which, if any, are dominant.\par
The data used for this work will be described in \textsection \ref{sec:data}. The results will be presented and explained in \textsection \ref{sec:results}. There will be a discussion of the results in \textsection \ref{sec:discussion}, and finally, our work will be summarised in a conclusion in \textsection \ref{sec:conclusion}. Throughout this work we adopt the following cosmological parameters: $H_0=\SI{72}{\km\per\s\per\mpc}$, $\Omega_\Lambda=0.7$, $\Omega_\text{M}=0.3$. We assume a Kroupa IMF \citep{kroupa2001}.
 \begin{figure*}
\centering
\includegraphics[width=\textwidth]{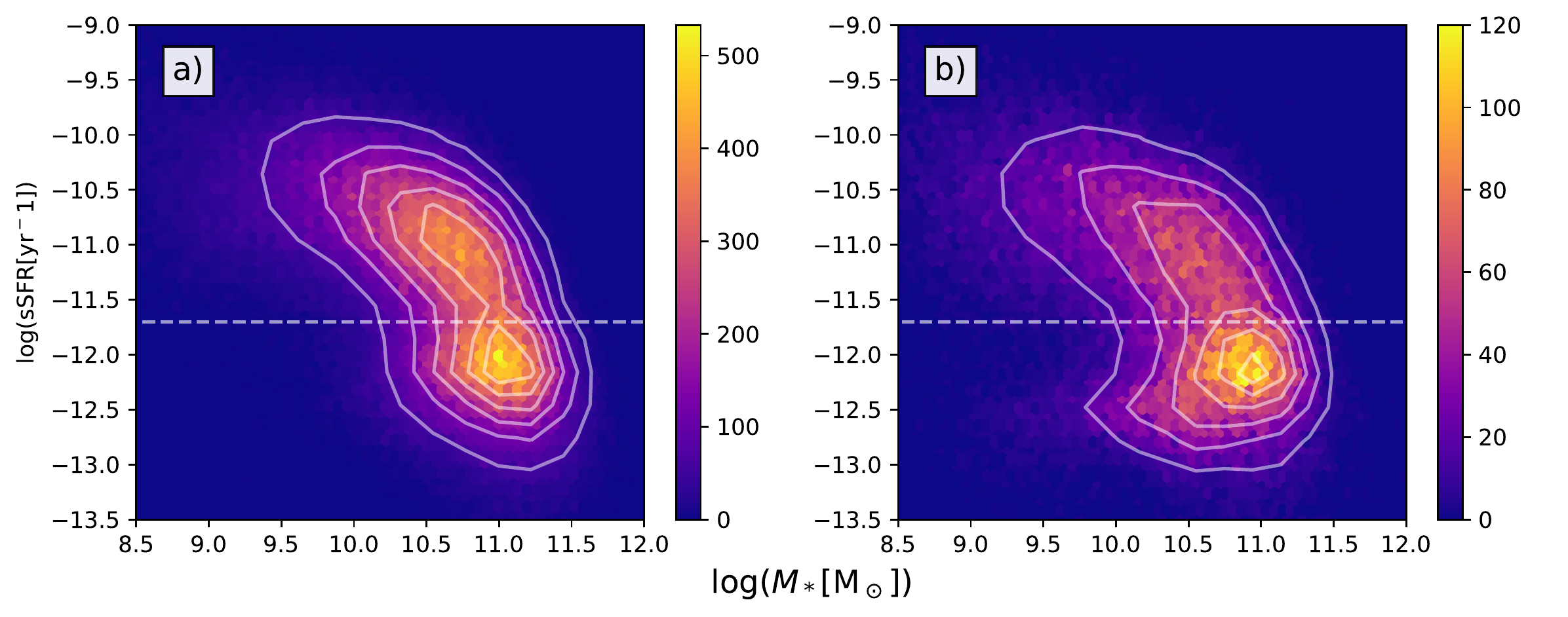}
\caption{Stellar mass ($M*$) versus specific star-formation rate ($\text{sSFR}_{\text{H}\upalpha}$) distributions for a) central galaxies and b) satellite galaxies. The white dashed line signifies a) sSFR$_{\text{H}\upalpha} = 10^{-11.7}$ yr$^{-1}$ and b) sSFR$_\text{UV} = 10^{-11.1}$ yr${-1}$. The $z$-axis denotes number of galaxies in each bin, with contour lines indicating regions of equal numbers. The dark blue regions signify no data.}
\label{fig:ha_cs}
\end{figure*}

\section{Data}\label{sec:data}
Our sample is derived from the set of galaxies with spectroscopic measurements in the Sloan Digital Sky Survey (SDSS) Data Release 7 with $0.01<z\leq0.2$ \citep{sdssdr7}. Although more recent data releases are available from SDSS, this version contains several `value-added' catalogues which we use. For each galaxy, we use the survey completeness correction given by \citet{blanton2005} to mimic a complete, flux-limited sample to the depth of the SDSS spectroscopic catalogs. Further, we use the volume corrections calculated by \citet{omand2015}\footnote{\url{http://cdsarc.u-strasbg.fr/viz-bin/cat/J/MNRAS/440/843}} to ensure selection effects do not significantly bias our results. For each galaxy, we compile the stellar mass ($M_*$ [M$_\odot$]), H$\upalpha$ flux ([erg s$^{-1}$ cm$^{-2}$]) and equivalent width (EW [\AA]) from the MPA-JHU value added catalogs \citep{brinchmann2004}\footnote{\url{https://wwwmpa.mpa-garching.mpg.de/SDSS/DR7/}}. \par

The UV emission measurements are taken from the GALEX-SDSS-WISE Legacy Catalog Medium-Deep 2 (GSWLC-M2) \citep{galex2005,galex2007,galex2016,galex2018}\footnote{\url{http://pages.iu.edu/~salims/gswlc/}}\footnote{\url{http://galex.stsci.edu/GR6/}}. For our purposes, this catalogue provides well-matched GALEX photometry to the SDSS spectroscopic galaxy sample. The GALEX satellite provided photometry in both the Near Ultraviolet (NUV; 2300 \AA) and Far Ultraviolet (FUV; 1500 \AA). Although the FUV is a cleaner probe of the recent star-formation rate in local galaxies, there are many more measurements of galaxies in the NUV due to a failure of the FUV instrument. We will use NUV as our ultraviolet probe in this paper, but we note that we have explored using FUV instead and find no qualitative change in our results except for a reduction in the statistical power. After matching our galaxy sample and its associated stellar mass and H$\upalpha$ measurements to the GSWLC-M2, and removing galaxies with flagged photometry or non-detections (e.g. flux in H$\upalpha$ or UV < 0), we are left with a sample of 227,176 galaxies with H$\upalpha$ and UV coverage. Note that the fraction of removed galaxies due to poor NUV observations is about 2 per cent across all stellar mass bins, with a slight increase towards the highest mass bins. \par

The GALEX observations which go into the GSWLC-M2 catalog have variable exposure times across the different pointings in the range of 650 to 4000 seconds in the NUV band \citep{galex2018}. Thus the flux limit also varies across these pointings. However, within the footprint of the GALEX observations, essentially all SDSS galaxies with $r < 16$ are detected. For blue galaxies ($u-r< 0.8$), this universal detection remains until past the $r$ = 17.7 limit of the SDSS main spectroscopic sample. For red galaxies ($u-r>0.8$), 75 per cent of them are detected at $r$ = 17, while about 50 per cent are detected at 17.7. Thus, our catalog is missing some red SDSS spectroscopic galaxies at the limit of the survey. To overcome this, when we rely on detections of low star-formation rates in UV, we reduce the redshift limit of the sample to z = 0.1, where we are nearly complete.  \par 
  
Distances are calculated assuming the redshift is dominated by the Hubble flow, and these distances are used to compute luminosities from the fluxes. Star-formation rates (SFR [M$_\odot$ yr$^{-1}]$) and specific star-formation rates (sSFR [yr$^{-1}$]) are computed using H$\upalpha$ and NUV flux calibrations from \citet{kennicutt1998} as in Equations \ref{eqn:halpha} and \ref{eqn:uv}:
\begin{equation}\label{eqn:halpha}
    \text{SFR}_{\text{H}\upalpha} = 7.9\times 10^{-42} L_{\text{H}\upalpha} [\text{erg s}^{-1}],
\end{equation}
\begin{equation}\label{eqn:uv}
    \text{SFR}_{\text{UV}} = 1.4\times 10^{-28} L_\nu [\text{erg s}^{-1} \text{Hz}^{-1}].
\end{equation}
It should be noted that our results are not sensitive to the exact form of these calibrations as we are principally interested in the differential change between environments and our divisions between `star-forming' and `passive' are motivated by empirical divisions in the observed data. For all our results, the galaxies are weighted by the completeness and volume corrections from \citet{blanton2005} and \citet{omand2015}. The statistical errors are calculated using raw data.\par

We obtain 221,904 petrosian magnitudes across all SDSS bands $u, g, r, i, z$ \citep{doi2010} and match these to emission line data and stellar masses in order to produce a colour magnitude diagram with star-formation information. These magnitudes satisfy a redshift limit $z<0.2$ and magnitude limit $u<20$.

Our environmental (central/satellite) classifications come from the \citet{yang2007} group catalogue (hereafter YGC). In the YGC, galaxies were assigned to groups via the friends-of-friends algorithm and the groups were assigned a total mass by rank-ordering the total luminosity (or stellar mass) of the group members, for a given comoving volume and halo mass function. This rank-ordering ensures that the halo mass function for a given cosmology is consistent with the total masses in the groups. The most luminous or the most massive (in terms of stellar mass) galaxy in a group is considered the central galaxy, while the remaining galaxies are denoted as satellites. Again, note that this means ungrouped, or isolated, galaxies are by definition also central galaxies.  For this study, we assume a central galaxy is the one with the most stellar mass. \par

From the YGC, we obtain the satellite/central classification for each galaxy and the total mass of the halo ($M_h$) that the galaxy resides in.  For the galaxies which have appropriate H$\upalpha$ and UV coverage, we find a sample of 222,304 galaxies within the YGC.\par

For each group with redshift $z_g$, we can also calculate the group-centric halo radius $r_{180}$ (Equation \ref{eqn:r180}), defined where the dark matter halo has an overdensity of 180,
   \begin{equation}\label{eqn:r180}
     r_{180} = 1.26\text{ } h^{-1} \text{ Mpc } \Bigg(\frac{M_h}{10^{14}h^{-1}\text{ M}_\odot}\Bigg)^{1/3}(1+z_g)^{-1},
 \end{equation}
 and the effective line-of-sight velocity dispersion $\upsigma$ (Equation \ref{eqn:sigma}),
  \begin{equation}\label{eqn:sigma}
     \upsigma = 397.9 \text{ km s}^{-1} \Bigg(\frac{M_h}{10^{14}h^{-1}\text{ M}_\odot}\Bigg)^{0.3214}
 \end{equation}
 as in \citet{yang2007}. In this manner, we calculate the projected separation between each satellite galaxy and its central galaxy in terms of $r_{180}$ of the group. We can use the redshift offset between the satellite and the group to calculate the relative velocity difference,
 \begin{equation}
    \Delta v = c(z_g-z_s).
 \end{equation}
In analysis, this is normalized as $\Delta v/\upsigma$.
 \section{Results}\label{sec:results}
 \subsection{Quenched galaxies}\label{subsec:quenched}
Although it is well established that `quenched' or passive galaxies are strong functions of stellar mass and environment, we will first examine this same phenomenon in our sample. Our motivation is to establish the main trends, as well as examining their dependence on the star-formation tracer---in particular H$\upalpha$ and UV. In Figure \ref{fig:uv_ha_hexbin}, we present the distributions of galaxies with respect to their stellar mass ($M_*$) and specific star-formation rate ($\text{sSFR}$), with SFR calculated using H$\upalpha$ in the left panel and SFR calculated using UV in the right panel. $\text{sSFR}$ is calculated simply by dividing the star-formation rate by the current stellar mass and gives an indication of the star-formation the galaxy has with respect to its integrated past star-formation rate. 

There are some obvious differences in the two panels of the figure which show the impact of SFR being calculated via H$\upalpha$ or UV flux which we will discuss after taking each in turn. In the $\text{sSFR}_{\text{H}\upalpha}$ panel, there is a clear bimodality leading to two particular over-dense regions, one due to star-forming galaxies, and one due to quenched galaxies. Principally, by visual inspection, a dividing line between the two populations was chosen such that quenched galaxies have $\log(\text{sSFR}/\text{yr}^{-1})<-11.7$, and star-forming galaxies are those above this cut-off.  This cut-off is lower than some sSFR demarcations in the literature, such as \citet{wetzel2012} who find that the trough in sSFR between star-forming and quenched galaxies in found at sSFR$ = 10^{-11}$ yr$^{-1}$ when using H$\upalpha$-based star-formation rates. However, it is important to remember that our H$\upalpha$-based SFRs are not corrected for the effects of internal dust extinction.

In Appendix \ref{sec:dust}, we examine the relative effects of dust extinction on central galaxies and satellite galaxies, with respect to several properties of the galaxies (eg, in stellar mass bins, halo mass bins, radial and velocity offset bins). We conclude that there is no difference in the attenuation effects between centrals and satellites in any of these subsamples. Since applying this attenuation correction would have no appreciable difference on the differential effects of centrals and satellites and thus no effect on our subsequent analysis, we do not apply it. This choice allows us to use H$\upalpha$ measurements for galaxies in which there is no H$\upbeta$ measured, and maintains a clear connection to the massive stars responsible for H$\upalpha$ emission. We opt to retain a larger sample of galaxies with well-measured (SNR > 3) H$\upalpha$ (153,294 centrals, 37,811 satellites), compared to those with well-measured H$\upalpha$ and H$\upbeta$ (125,346 centrals, 29,318 satellites).

Turning our attention to the right panel of Figure \ref{fig:uv_ha_hexbin}, we can see the distribution of galaxies in the  $\text{sSFR}_\text{UV}$--stellar mass plane. It is clear that with this star-formation indicator, the bimodality is still apparent as in the H$\upalpha$ case. However, we note that the division in the bimodality is different to that in sSFR$_{\text{H}\upalpha}$, and still not exactly equal to the common cut-off point in literature. Thus, we demarcate star-forming and quenched galaxies using a cut-off of $\log(\text{sSFR}/\text{yr}^{-1})<-11.1$, chosen from the local minimum of the sSFR$_{\text{UV}}$ distribution of the sample. \par

We can now investigate how the sSFR distribution varies with environment. In Figure \ref{fig:ha_cs}, we present the galaxy distribution in the sSFR--stellar mass plane, where we measured the sSFR using  H$\upalpha$, for central galaxies (left panel) and satellite galaxies (right panel).  There are interesting, but expected, differences between the environmentally separated populations. It is clear that the quenched population appears more prominent with respect to the star-forming population at all stellar masses in the satellite galaxies than it does in the central galaxies. It is worth noting the relative lack of quenched galaxies with $M_*\lesssim 10^{10.5}$ in the central population, but their prominence in the satellite population.  

\begin{figure}
    \centering
    \includegraphics[width=\columnwidth]{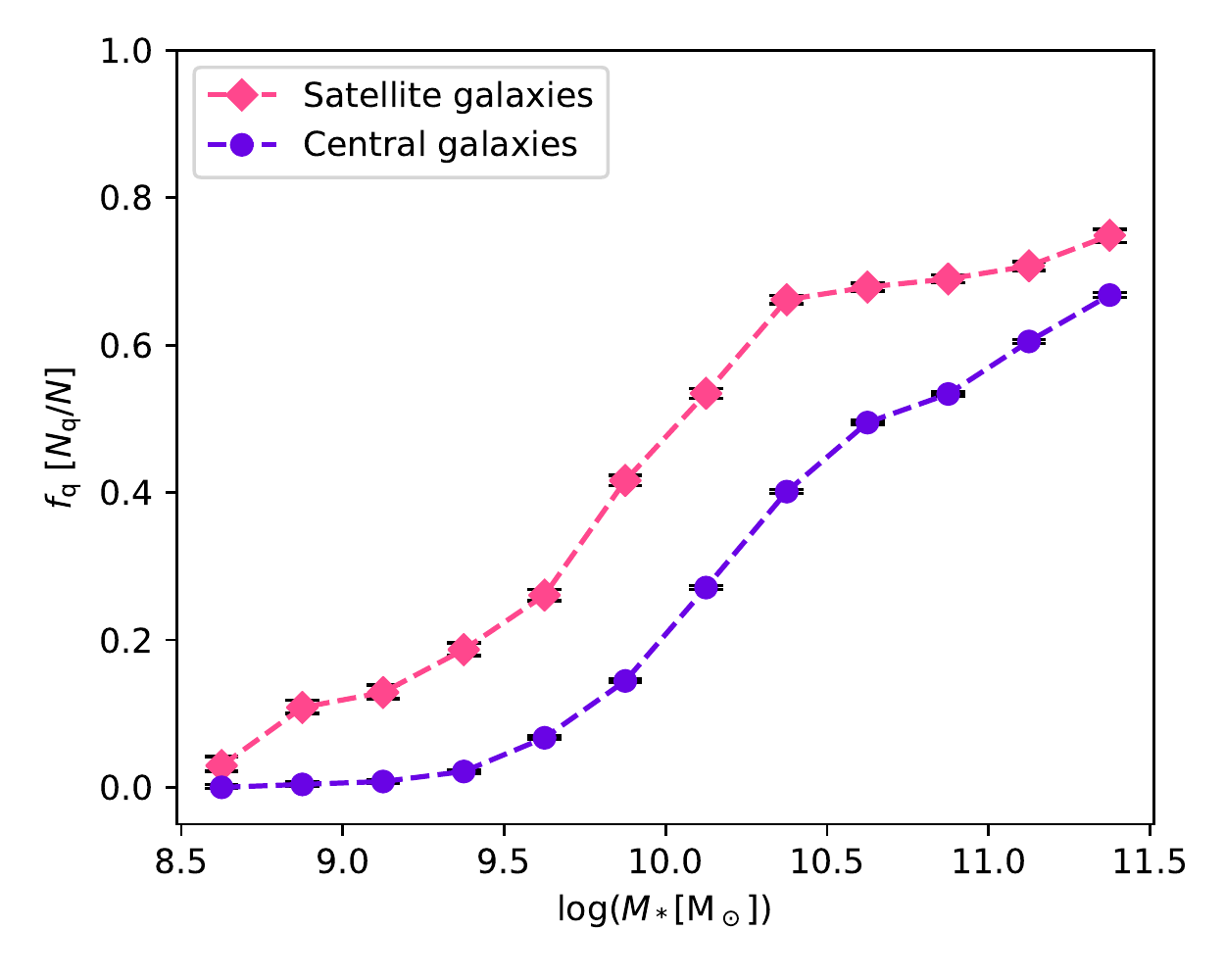}
    \caption{Fraction of quenched galaxies, $f_\text{q} = N_\text{q}/N$, where $N_\text{q}$ is the number of quenched central(satellite) galaxies and $N$ is the total number of central(satellite) galaxies. The star-formation rate is drawn from H$\upalpha$. }
    \label{fig:ha_frq}
\end{figure}

To have a more quantitative view of these distributions, we present Figure \ref{fig:ha_frq}, in which we bin the data in stellar mass, and calculate the fractions of quenched galaxies with respect to the total number of galaxies, $f_\text{q} = N_\text{q}/N$, for centrals and satellites separately. Error bars are determined using the percent point function of a beta distribution where the input parameters are the number of transient centrals (satellites) and the total number of centrals (satellites) in each mass bin \citep{cameron2011}. At all stellar masses, there is a larger fraction of quenched satellite galaxies than quenched central galaxies, with a sharp increase for both centrals and satellites at around $10^{10}$ M$_\odot$. This supports the assertion that galaxies in dense environments such as satellites, and particularly high stellar mass galaxies, are more likely to be quenched. This general shape is consistent with many other studies of central/satellite quenched fractions in the literature \citep{kimm2009, bluck14, davies2019}. Note that our procedure of removing undetected NUV galaxies means that these fractions may suffer some incompleteness at the massive end, and thereby underestimate the quenched fraction. This removal of galaxies accounts for up to 8 per cent of quenched galaxies at the high-mass end. We have checked that imposing a redshift constraint of $z \leq 0.1$ does not qualitatively change our results.
\begin{figure}
    \centering
    \includegraphics[width=\columnwidth]{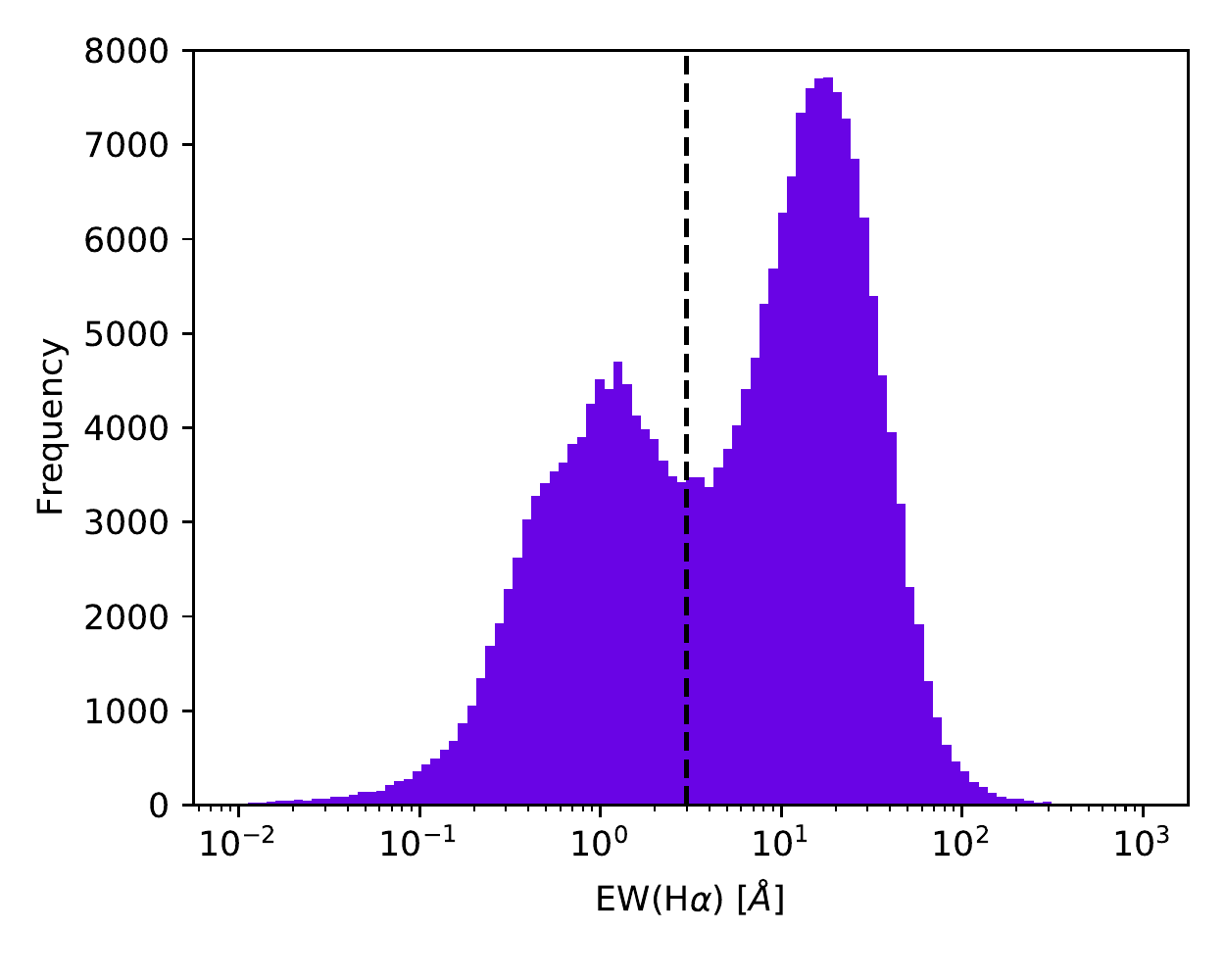}
    \caption{Distribution of equivalent widths of H$\upalpha$ measurements in angstroms. The black dashed line denotes EW(H$\upalpha)=3$\AA.}
    \label{fig:ew_hist}
\end{figure}
\begin{figure}
    \centering
    \includegraphics[width=\columnwidth]{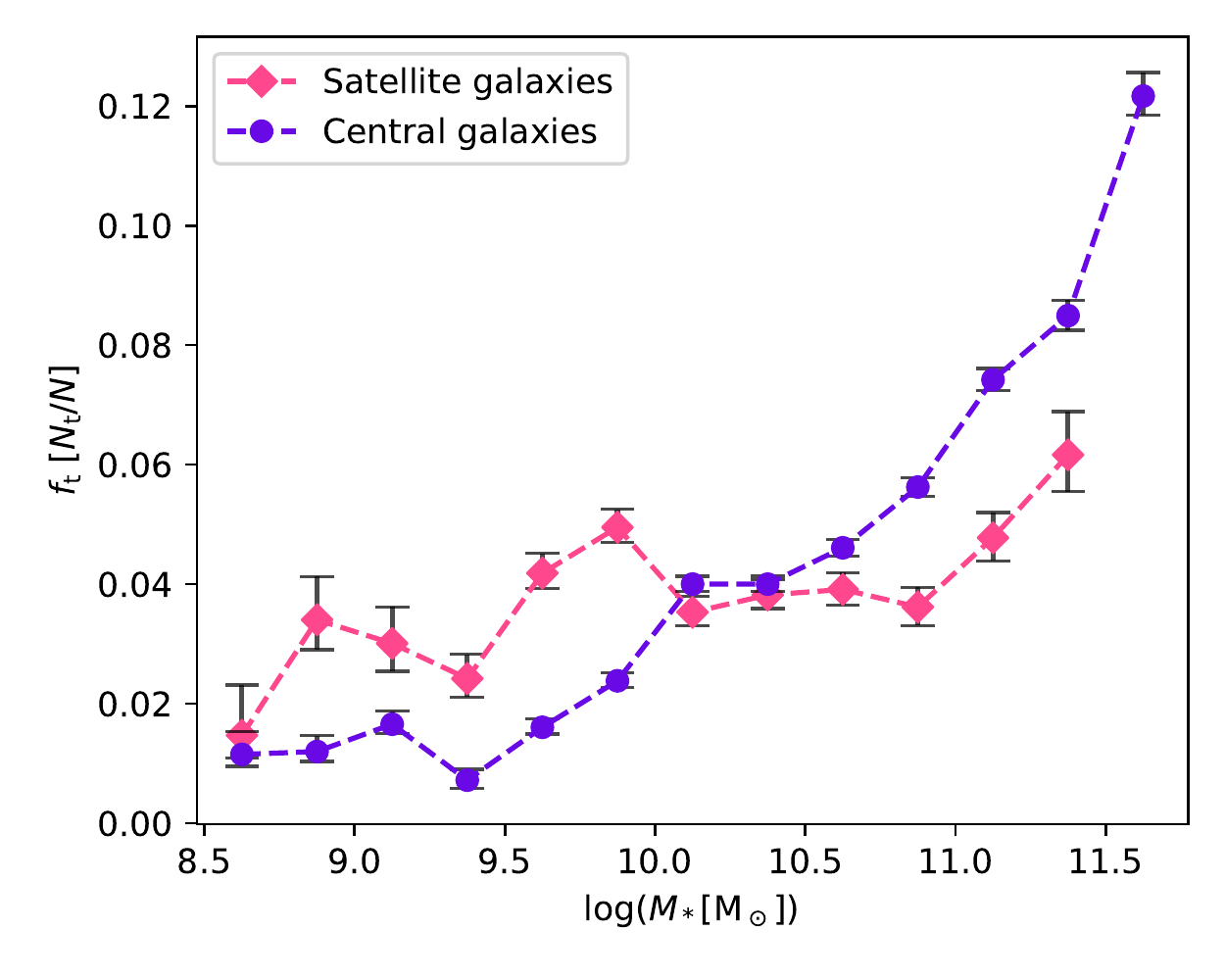}
    \caption{Fraction of transient galaxies, $f_\text{t} = N_\text{t}/N$, where $N_\text{t}$ is the number of transient central(satellite) galaxies and $N$ is the total number of central(satellite) galaxies. Transient galaxies are those with no detectable H$\upalpha$ and UV emission from star-formation.}
    \label{fig:sf2q_uv_frt}
\end{figure}
\begin{figure}
    \centering
    \includegraphics[width=\columnwidth]{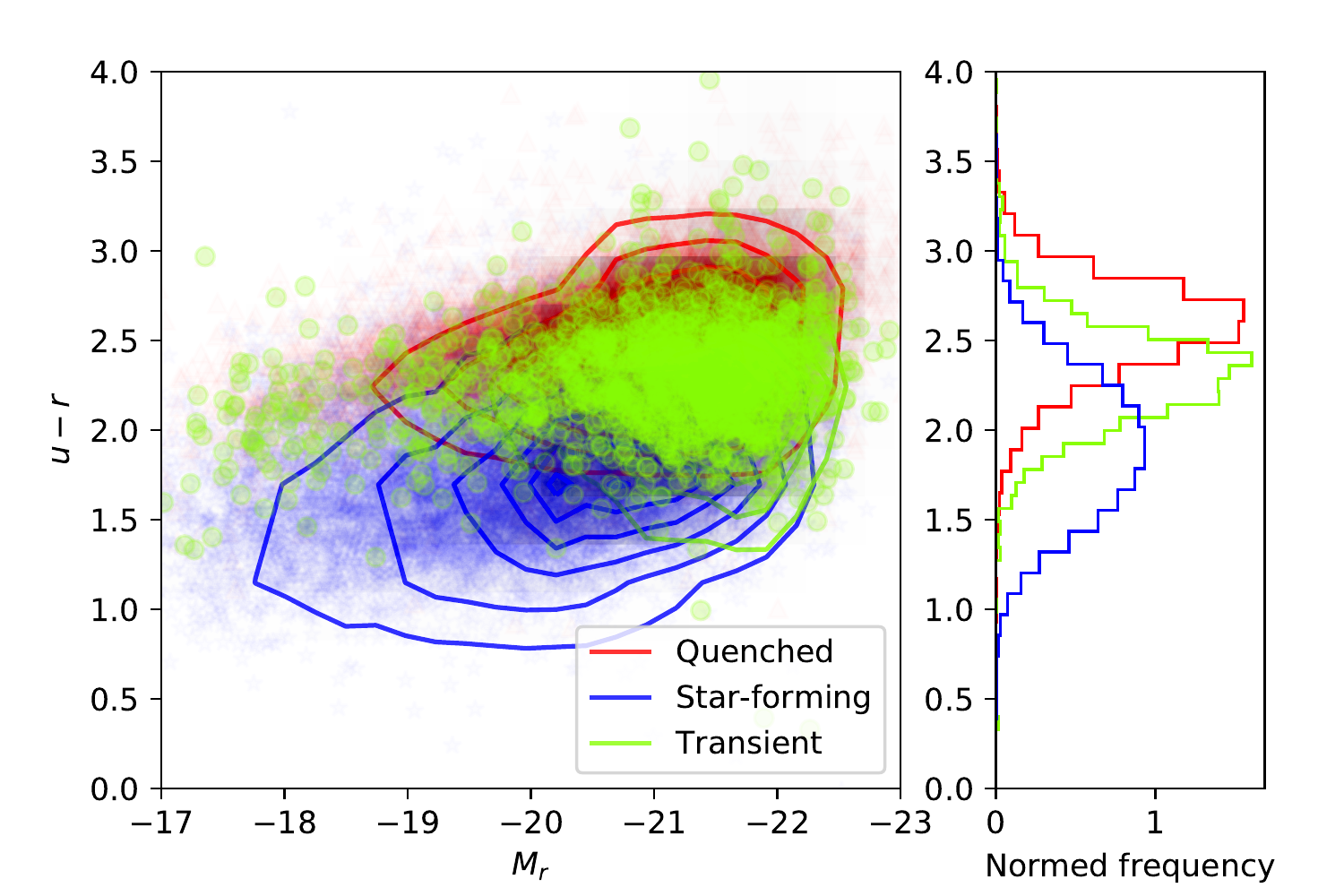}
    \caption{Colour magnitude diagram, $u-r$ against $M_r$. Contour plots indicate abundances of quenched and star-forming populations defined as above using sSFR$_{\text{H}\upalpha}$. A marginalised distribution shows that transient galaxies as defined in this paper lie between red and blue galaxies on a colour magnitude diagram. }
    \label{fig:cmd}
\end{figure}
\begin{figure*}
    \centering
    \includegraphics[width=0.9\textwidth]{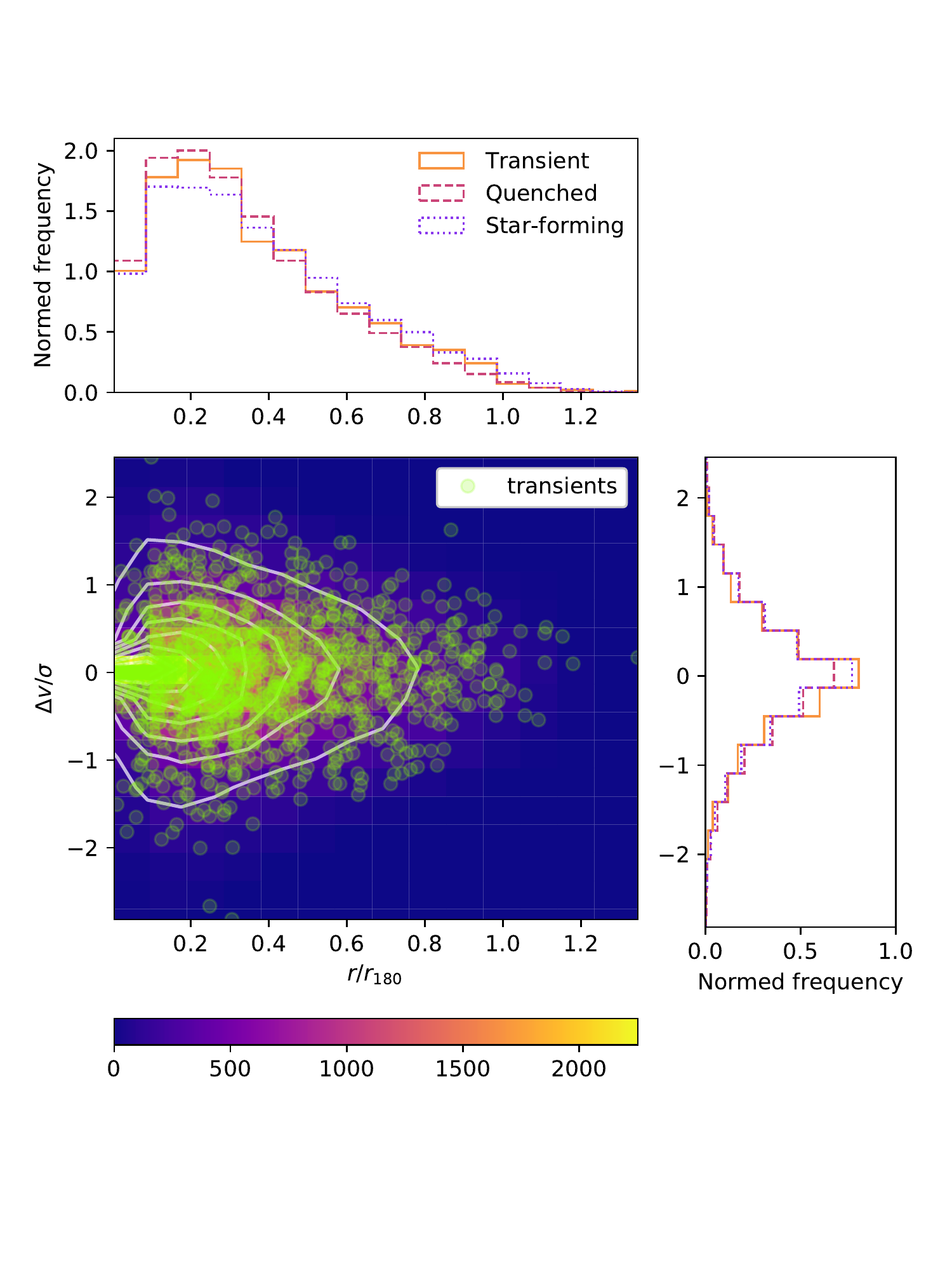}
    \caption{Phase space histogram showing $r/r_{180}$ and $\Delta v/\upsigma$ for all galaxies, with marginalised distributions on each axis for star-forming, quenched and transient galaxies. Note the similarity in distributions between transient, quenched, and star-forming galaxies. The $z$-axis denotes number of galaxies in each bin.}
    \label{fig:phase_diagram}
\end{figure*}
\begin{figure*}
    \centering
    \includegraphics[width=\textwidth]{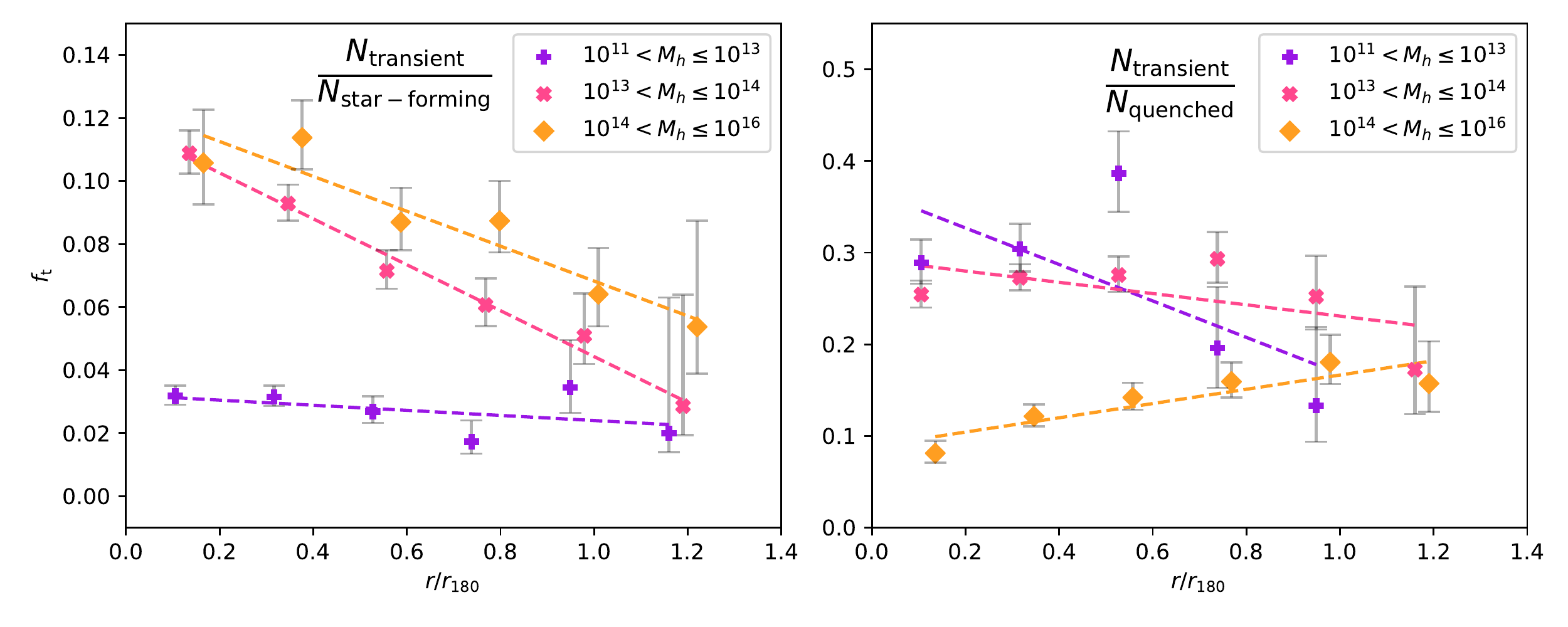}
    \caption{Fractions of transient galaxies with respect to star-forming galaxies (left) and quenched galaxies (right) for different group halo mass bins. There is a clear downward trend for the fraction of transient galaxies with respect to star-forming galaxies as $r/r_{180}$ increases at high halo mass.}
    \label{fig:fsfqr}
\end{figure*}

\begin{figure}
    \centering
    \includegraphics[width=\columnwidth]{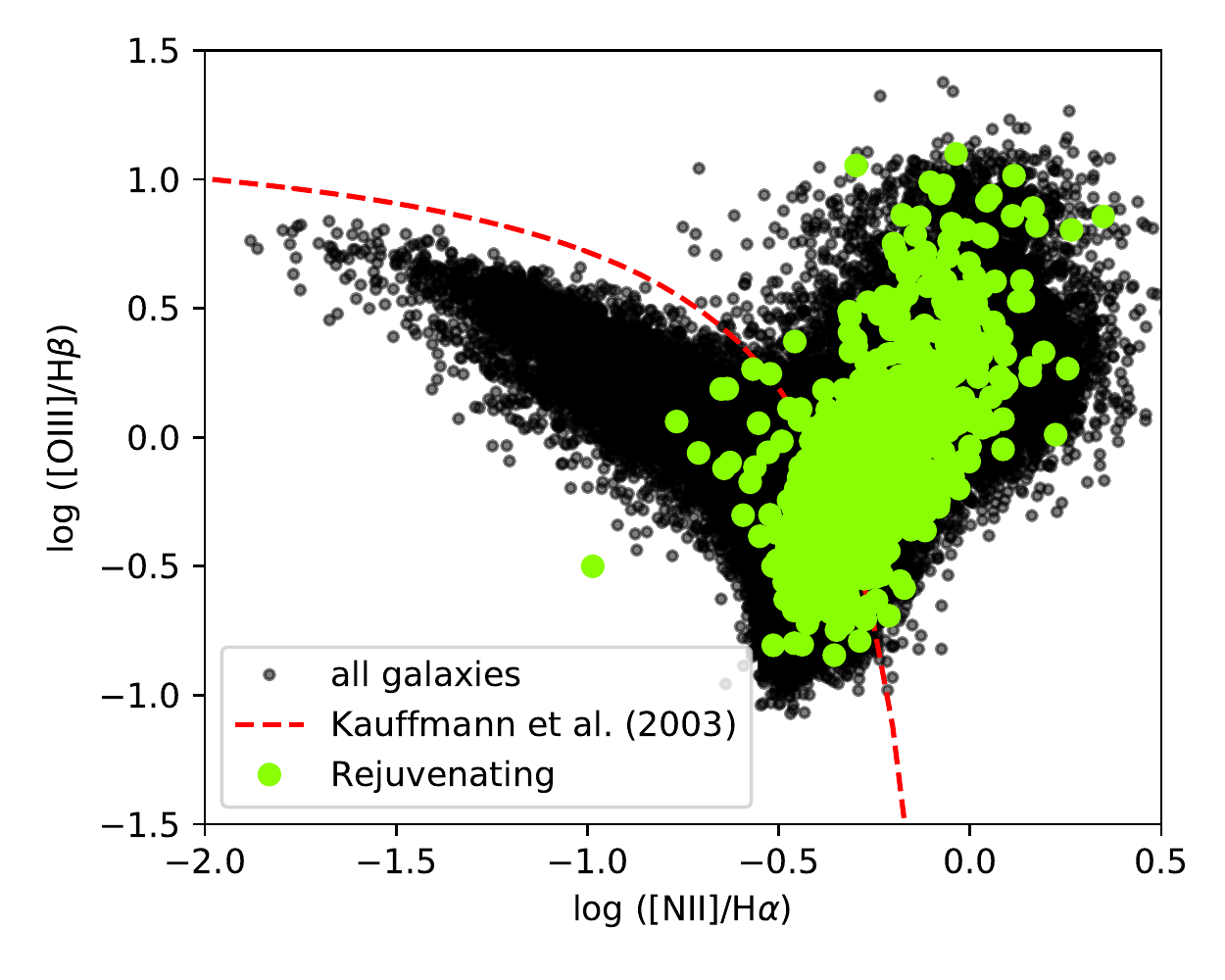}
    \caption{BPT diagram with the \citet{kauffmann2003} demarcation line. Two-thirds of the galaxies identified by our rejuvenating candidate selection are shown here to not have truly star-forming emission.}
    \label{fig:bpt}
\end{figure}

\begin{figure}
    \centering
    \includegraphics[width=\columnwidth]{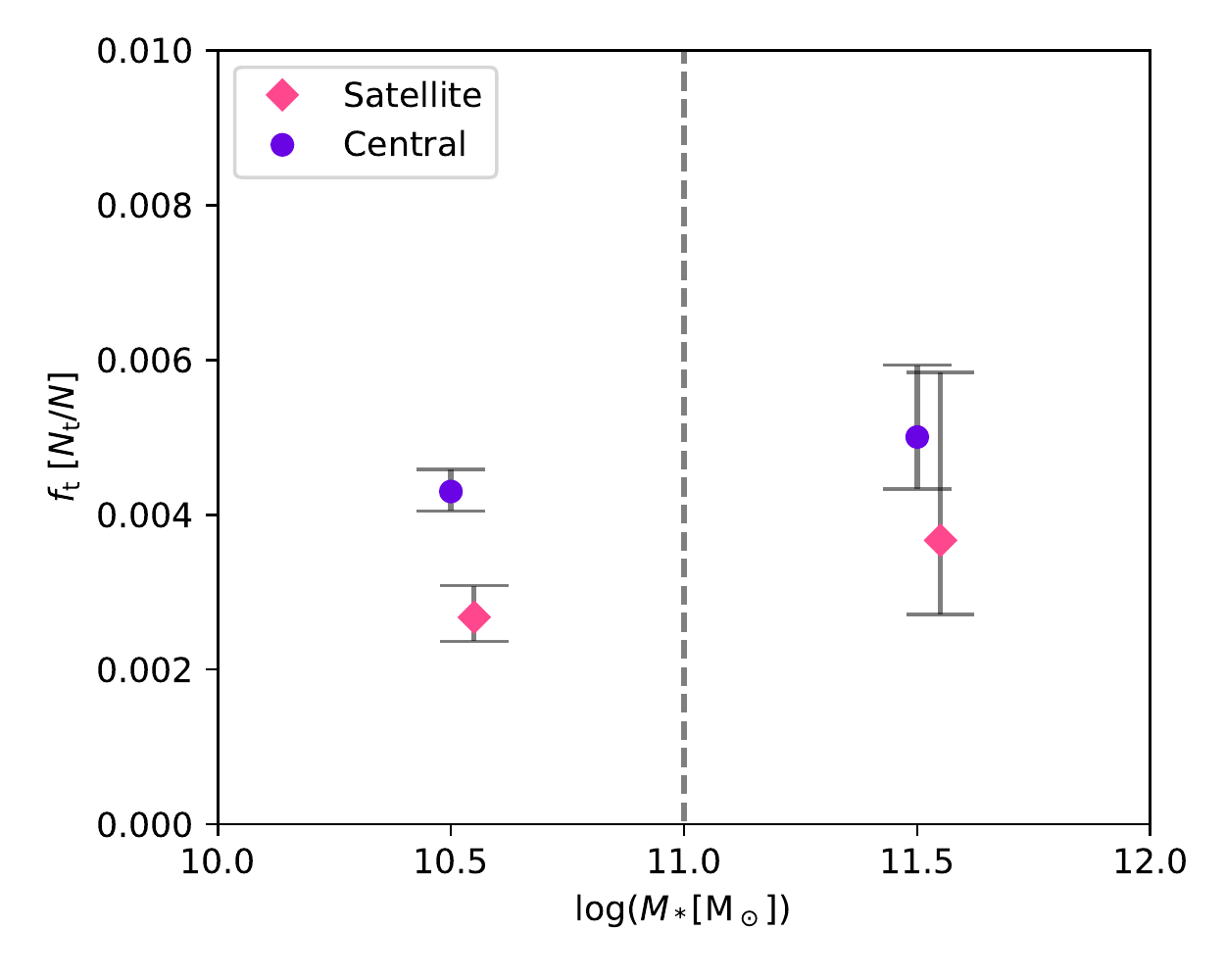}
    \caption{Fraction of candidate rejuvenating galaxies after removal of AGN-dominated galaxies, $f_\text{r} = N_\text{r}/N$, where $N_\text{r}$ is the number of candidate rejuvenating central(satellite) galaxies and $N$ is the total number of central(satellite) galaxies. The data is binned by stellar-mass, with low stellar-mass on the left and high stellar-mass on the right.}
    \label{fig:q2sf_uv_frt}
\end{figure}

\subsection{Transient galaxies}\label{subsec:transient}
Now that we have established the clear environmental trends in star-formation properties, and the implied quenching effect on satellite galaxies, we can attempt to pick out those galaxies which have recently undergone the cessation of their star-formation. As discussed in the introduction, H$\upalpha$ and UV measurements are sensitive to star-formation on different timescales (10 Myr and 100 Myr respectively), so a combination of the two measurements can be applied to investigate the differential star-formation rate. Consider a galaxy which has a star-formation rate typical of a star-forming galaxy. If this star-formation is shut off quickly in a semi-instantaneous event, the O stars giving rise to the H$\upalpha$ emission will die within 10 Myr and the galaxy will cease to appear star-forming in this tracer. However, as the UV emission comes from the less massive stars, the galaxy will appear to have significant UV emission until those stars die $\sim$ 100 Myrs after the extinction event. Thus, we can look for these `transient' galaxies by isolating those that have a star-forming UV component, but a negligible H$\upalpha$ component. 

To define these `transient' galaxies we must ensure that they have H$\upalpha$ flux consistent with no star-formation, and that the UV component must be from star-formation and not residual emission from old stars, and that this UV emission is significant. First, to determine the galaxies with a lack of H$\upalpha$ emission directly, we plot a histogram of the distribution of the equivalent width (EW) of H$\upalpha$ in Figure \ref{fig:ew_hist}. There are two clear populations: galaxies with a significant H$\upalpha$ detection, and galaxies with negligible H$\upalpha$ detection with a clear dividing line at EW(H$\upalpha) = 3$\AA. Galaxies below this cut are assumed to have no significant H$\upalpha$ emission.\par

Similarly, we want to establish a $\text{sSFR}_{\text{UV}}$ cut which is consistent with the UV emission being the normal emission of a star-forming galaxy. For this, we will return to our $\log(\text{sSFR}_{\text{UV}}/\text{yr}^{-1}) > -11.1$ cut established in the previous subsection. However, we further require that the galaxy has a $\text{NUV}-r$ colour consistent with being a star-forming galaxy. From an analysis of the colour magnitude diagram, and consistent with previous divisions, we require $\text{NUV}-r\leq4.5$ to be star-forming. \par

We can now define transient galaxies as those with $\log(\text{sSFR}_{\text{UV}}/\text{yr}^{-1}) > -11.1$, $\text{NUV}-r\leq4.5$ and $\text{EW}(\text{H}_\upalpha) < 3$\AA.  In Figure \ref{fig:sf2q_uv_frt}, the fractions of transient galaxies with respect to the total number of galaxies ($f_\text{t}=N_\text{t}/N$) for centrals and satellites are plotted for each stellar mass bin. The clearest difference between the two distributions is the excess of transient fraction in satellites compared to centrals at low $M_*$, compared to the lack thereof at higher stellar masses. Over the entire low mass bin ($M_*<10^{10.5}$ M$_\odot$), there is an average 1 percentage point excess in transient fraction of satellites compared to centrals at a $6.3\upsigma$ significance.

Across all stellar mass bins, we count 2,758 transient satellite galaxies and 12,080 transient central galaxies. This gives 5.9 per cent and 6.7 per cent fractions of the total satellite/central galaxy count respectively. However, we find further evidence for a mass dependence by looking only at galaxies with $M_* < 10^{10}$ M$_\odot$, where we find 303 transient satellite galaxies and 340 transient central galaxies, giving 2.7 per cent and 1.2 per cent of the total number of satellite/central galaxies with $M_* < 10^{10}$ M$_\odot$ respectively. We see a slight increase in the fraction of transient satellite galaxies at low-mass versus transient central galaxies. This implies that although high-mass galaxies are more likely to be quenched, satellite transient galaxies deviate from transient centrals at low masses, meaning low-mass satellites may be more inclined to begin to quench based on environmental factors. 

We plot a colour magnitude diagram (CMD) in Figure \ref{fig:cmd}. Contour lines indicate populations of quenched and star-forming galaxies, using definitions established in \textsection \ref{subsec:quenched}, using sSFR$_{\text{H}\upalpha}$. Individual transient galaxies are plotted over the contours. A histogram is plotted showing the distributions of colours of the red and blue galaxies. The CMD and histogram confirm that transient galaxies, as defined in this paper, lie between red and blue galaxies in terms of colour, and thus in terms of their evolution.

\begin{figure*}
    \centering
    \includegraphics[width=\textwidth]{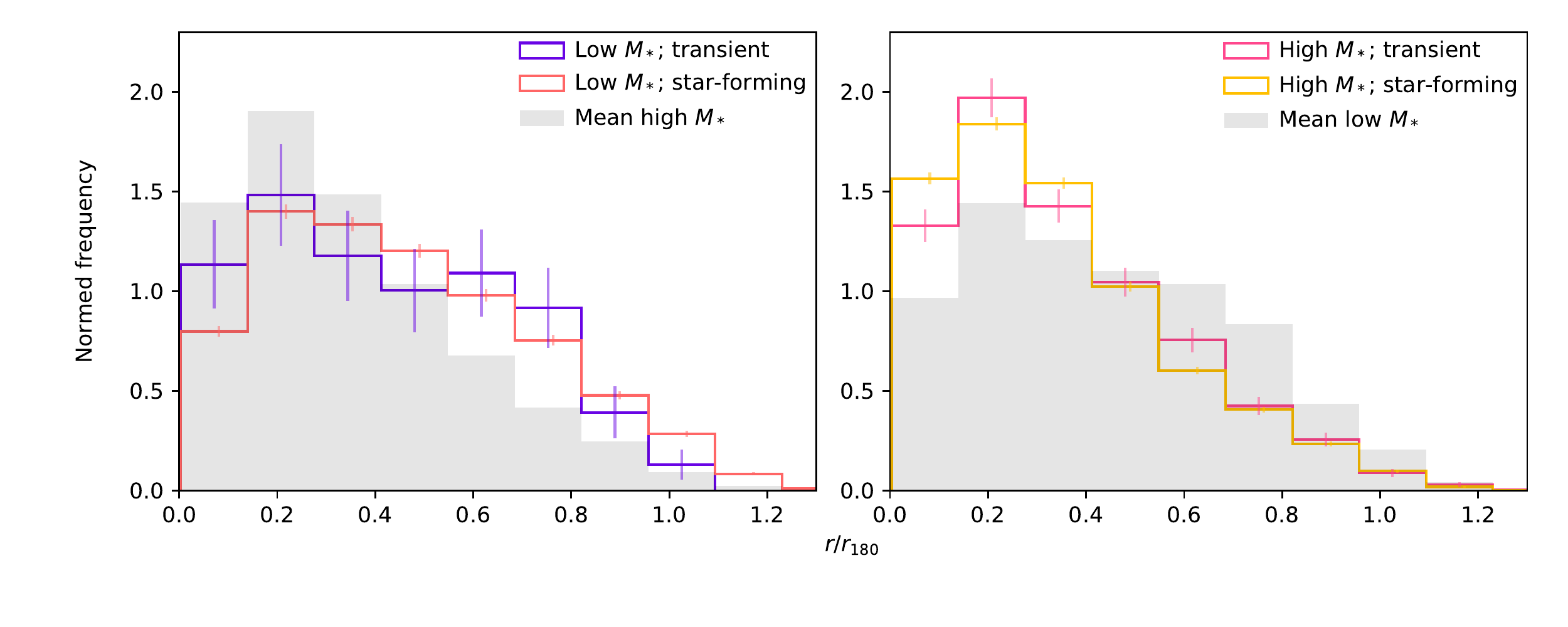}
    \caption{Histograms of stellar mass binned transient and star-forming galaxies with respect to $r/r_{180}$. Here, low stellar mass (left panel) refers to galaxies with stellar mass $10^6<M_*\leq10^{10}$ M$_\odot$ and high stellar mass (right panel) refers to galaxies with stellar mass $10^{10}<M_*\leq10^{13}$ M$_\odot$. The mean distribution of each plot is shown for ease of comparison. Errors are computed using the normalised count in each bin. }
    \label{fig:t_lmhm}
\end{figure*}
\begin{figure*}
    \centering
    \includegraphics[width=\textwidth]{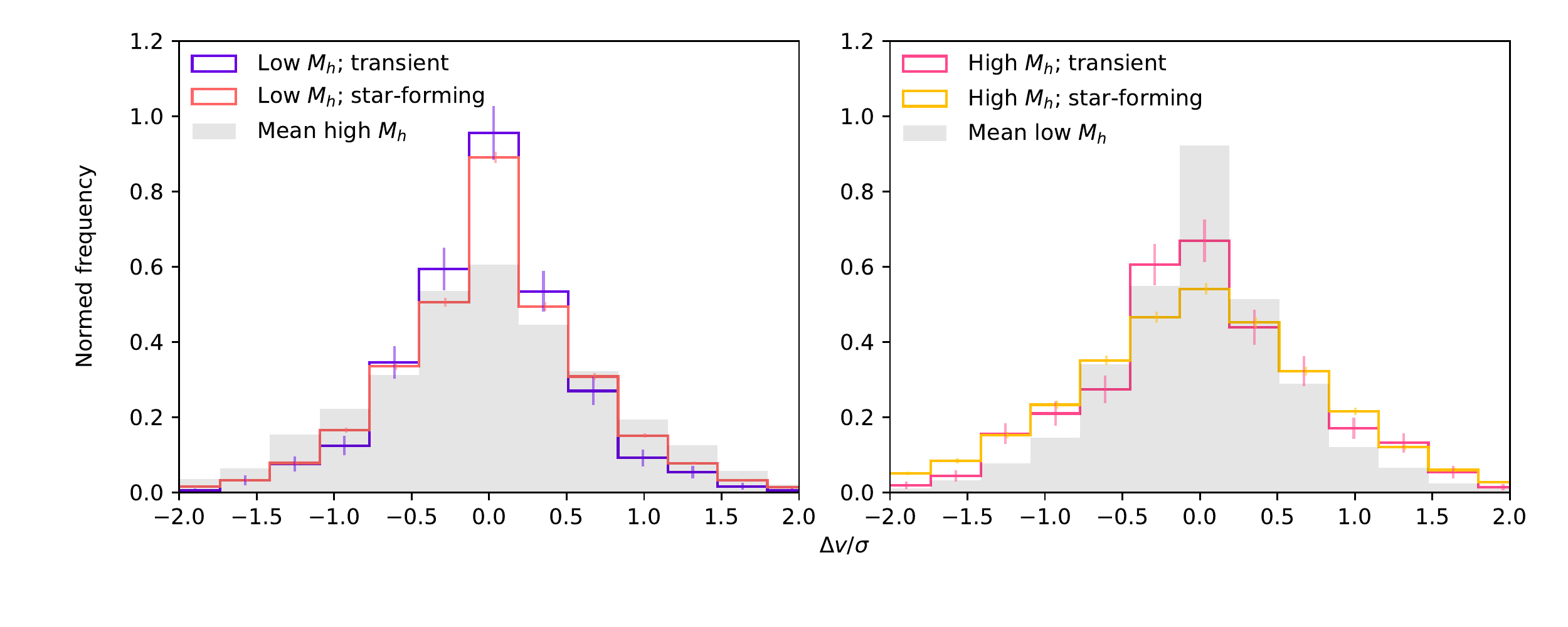}
    \caption{Histograms of halo mass binned transient and star-forming galaxies with respect to $\Delta v/\upsigma$. Here, low halo mass (left panel) refers to groups with halo mass $10^{11}<M_h\leq10^{13.5}$ M$_\odot$ and high halo mass (right panel) refers to groups with halo mass $10^{13.5}<M_h\leq10^{16}$ M$_\odot$. The mean distribution of each plot is shown for ease of comparison. Errors are computed using the normalised count in each bin. }
    \label{fig:t_halomass_delv}
\end{figure*}
\subsection{Group and satellite properties} \label{sec:group_satellite_prop}
Now that we have identified a population of galaxies with recently quenched star-formation and which are preferentially satellite galaxies, we can investigate the environments of those galaxies in more detail. The phase space distribution of galaxies in groups or clusters, that is, the plane of projected group-centric distance versus the relative velocity offset, has shown to hold important information on the path of a satellite upon entering the system \citep{muzzin2014,jaffe2018}. For each satellite, we calculate their relative velocity ($\Delta v$) with respect to the group velocity dispersion ($\upsigma$) and the projected separation between the satellite and the central ($r$) with respect to the group halo radius ($r_{180}$). Figure \ref{fig:phase_diagram} shows a 2d histogram with axes $r/r_{180}$ and $\Delta v/\upsigma$, which shows the distribution of the entire sample of satellite galaxies across the phase space with marginalised distributions on each axis for star-forming, quenched, and transient satellite galaxies. Individual transient galaxies are indicated on the plot by green circles. The distributions appear similar across the three distributions and in both the velocity distribution and radial distribution. There is a slight suggestion that the transient population has a radial distribution more similar to the quenched population than the star-forming population, however KS-tests on all the distributions show no significant trends. However, this plot includes all satellites regardless of their stellar mass. We will investigate any possible stellar mass dependence further in \textsection \ref{sec:discussion}.\par

We examine the excess of satellite transients further by looking at the relative proportions of transient satellites with respect to star-forming galaxies and quenched galaxies binned by group halo mass. In Figure \ref{fig:fsfqr}, we plot the fraction of transient galaxies compared to star-forming galaxies against the radial position of the satellite galaxy in the group (left panel). We also plot the fraction of transient galaxy compared to quenched galaxies (right panel). First, examining the panel presenting the fraction of transients with respect to the star-forming galaxies, we see two interesting trends. Firstly, there is a clear radial trend in that this ratio increases towards the center of the halo. This would be expected if satellites become transients upon infalling into the group, and that their likelihood of quenching increases as they reach the center of the cluster. Secondly, there is a halo mass dependence to the transient/star-forming ratio. Halo masses greater than 10$^{13}$ M$_\odot$ host much higher rates of transients than do halos less massive. Again, this may indicate that transients occur in satellites after infalling, and that their rate of going through a transient stage is related to the densities of gas or size of gravitational potentials they encounter. 

In the right panel of Figure \ref{fig:fsfqr} the fraction of transient to quenched galaxies is shown for the same bins of halo mass and group-centric radius. For massive halos ($>$ 10$^{13}$ M$_\odot$), there is a clear opposite radial trend, in that this ratio is smaller with decreasing radius. Since the quenched galaxies are a larger proportion of all galaxies towards the group centre, this means that the transient population is anti-correlated with the quenched population. Usefully, this means that transients are not likely to be a momentary, recurrent phase in the life of quenched galaxies, but rather indicate a once-off event. Interestingly, the halo mass dependence of this ratio is less clear than it was in the transient/star-forming ratio and may suggest that the past level of quenching is not as closely correlated with current quenching. If quenching in a group occurs on short timescales with respect to the formation of the group, then this relative lack of correlation would be expected.  

\subsection{Rejuvenating galaxies}
The anti-correlation between the transient population and the quenched population suggests that the transients we are seeing are not a stage in the life of a normal quenched galaxies, but rather a phase indicative of a key quenching event. However, it is worth noting that galaxy-group and galaxy-galaxy interactions or continued further gas accretion may cause star-formation to begin again in a quenched galaxy, so called `rejuvenation'. Indeed, models of galaxy formation in which the gas accretion (and subsequent star formation is closely tied to the dark matter accretion suggest that galaxies should naturally go through periods of quenching and rejuvenation. For instance, a model which ties gas accretion to dark matter accretion finds that between 20 per cent and 80 per cent of galaxies at $z =0$ (depending on their stellar mass) go through a rejuvenation phase of at least 300 Myr of quiescence, followed by 300 Myr of star formation \citep{behroozi2019}. This motivates a short-timescale search for galaxies that are just beginning their rejuvenation. 

For this reason, in this paper, rejuvenating galaxies are defined as those with significant H$\upalpha$ detection [EW(H$\upalpha) \geq 3$\AA] but negligible UV emission, i.e. sSFR$_\text{UV} <10^{-11.1}$ yr$^{-1}$. Using similar logic in defining transient galaxies, this implies that star-formation has recently begun again in these galaxies. Due to requiring sensitivity to low-levels of star-formation from the UV emission, we limit our analysis in this section to galaxies at $z<0.1$. Note that this is a shorter-timescale rejuvenation selection than previous studies \citep[see, e.g.,][for low- and high-redshift analyses respectively]{thomas2010,chauke2019} which used spectral features to determine if galaxies had ever gone through a rejuvination phase. The spectral features are only sensitive to longer timescales ($>$100 Myr) than ours. Our selection is designed to find those galaxies just in the process of rejuvenating, and are not the integral constraint on the fraction that previous studies are. 

We stress, that while our selection of galaxies may include the sought after short-timescale rejuvenating galaxies, it is not exclusive. It will contain contaminants and an independent verification of the true nature of these potentially rejuvenating galaxies is difficult to obtain since most photometric and spectroscopic indicators vary on longer timescales than is relevant. Thus, we will remove galaxies from the sample which are not obviously of rejuvenating origin, but it should be cautioned that what is left is not necessarily rejuvenating.

The largest source of contamination will be AGN. Although optical line AGN are relatively rare in galaxies without significant NUV emission \citep{kauffmann2007}, we will need to take this into account. In Figure \ref{fig:bpt}, we show the BPT diagram for the whole sample of galaxies (black dots) and for those passing our rejuvenation selection (green dots). We also show the demarcation line of \citet{kauffmann2003} which divides purely star-forming galaxies (below the line) from composite or AGN-dominated galaxies (above the line). By this stringent definition, only 1/3 of our rejuvenation candidate galaxies have emission lines caused by pure star formation. 

Removing the AGN-dominated galaxies from the sample does not necessarily leave only rejuvenating galaxies, but it is now worth examining the stellar mass and environmental characteristics of these candidates.
In Figure \ref{fig:q2sf_uv_frt}, we show the ratio of the number of rejuvenating galaxy candidates over the total number of galaxies as a function of two stellar mass bins, low stellar mass ($10^{10}\leq M_*/$M$_\odot < 10^{11}$) and high stellar mass ($10^{11}\leq M_*/$M$_\odot < 10^{12}$) for each of the central and satellite populations. Note the minimum stellar mass of the rejuvenating galaxy candidates does not go below $10^{10} M_*$. The fraction of rejuvenating galaxy candidates in the low stellar mass bin is about 0.2-0.4 per cent, while it is about 0.4-0.5 per cent in the higher stellar mass bins. This mild stellar mass dependence is at least partially due to the underlying stellar mass dependence of the quenched fractions, as it is not possible to have a rejuvenated galaxy without first quenching it. Interestingly, there is only a mild difference in the fraction of rejuvenated galaxy candidates when divided into central and satellite galaxies. Quantitatively, the slight excesses of rejuvenating candidate centrals seen in the left and right panels of Figure \ref{fig:q2sf_uv_frt} have significances of 3.6$\upsigma$ and 0.6$\upsigma$ respectively. Thus, the data does not strongly indicate a significant excess of rejuvenating candidate centrals compared to satellites. In terms of bulk numbers, We find 257/59,784 rejuvenating candidates in low stellar mass central galaxies, and 39/7,795 rejuvenating candidates in high stellar mass central galaxies. This compares to 55/20,565 rejuvenating candidates low stellar mass satellite galaxies and 6/1,636 rejuvenating high stellar mass satellite galaxies. The behaviour in this plot is in contrast to the peak in transient satellite galaxies compared to central galaxies at $M_*\sim10^{9.5}$ M$_\odot$ in Figure \ref{fig:sf2q_uv_frt}. While we see clear evidence that the transients are environmentally effected, the rejuvenated candidate galaxies are not. This gives confidence that our transient population is a true population of galaxies undergoing a catastrophic quenching event.

For galaxies with stellar mass $M_*>10^{10}$ M$_\odot$, the fraction of rejuvenating candidate galaxies which have their emission due purely to star formation is 0.5 per cent. If these galaxies are all truly rejuvenating and the timescale with which we would detect them is $\sim 50$ Myr, then we would expect 10 per cent of galaxies to go through this phase every Gyr. This number is not incompatible with the results from galaxy formation models, or the other methods of measuring longer timescale rejuvenation. It is a matter for future work to make a holistic confrontation of short- and long-timescale rejuvenation with galaxy formation models.

\section{Discussion}\label{sec:discussion}

These results show that there are multiple aspects at play which influence the star-formation of galaxies and which depend on various galactic properties. We see enhanced quenching in satellites, likely due to the environmental effects that act on a satellite galaxy as it moves through a group. To get a clearer picture of how the environment effects a satellite galaxy, we can isolate galaxies that are in the process of quenching, and relate the individual galactic properties with those of the group. We note that there is a higher fraction of transient satellite galaxies at $M_*<10^{10}$ M$_\odot$ compared to transient central galaxies. This suggests that low-mass satellite galaxies are quenched more efficiently than low-mass centrals are. This scenario of low-mass satellite quenching is consistent with previous studies using direct stellar ages \citep{haines2006, smith2009}. It also shows similar trends to studies of spectroscopically selected post-starburst galaxies \citep{paccagnella2017}, although with a different selection method from ours.

It was found in \textsection \ref{sec:group_satellite_prop} that transient galaxies evenly occupy the full range of $r/r_{180}$, and that there was not a significant difference in the distributions of transients, star-forming and quenched galaxies.  This is, perhaps, inconsistent with \citet{muzzin2014} who find a particular ring structure in the phase-space location of z $\sim$ 1 post-starburst cluster galaxies. Similarly, \citet{owers2019} found that H$\updelta$ strong galaxies are principally found at small ($<$ 0.5 R$_{200}$) cluster-centric distances.  However, it is clear that the exact definition of a `quenching' galaxy, and the stellar mass limits will make a difference \citep{paccagnella2017}. Our result was shown for the full range of stellar masses, and we have seen that the most prominent differences may occur at low stellar mass. For this reason, we separate low-mass and high-mass satellites. In Figure \ref{fig:t_lmhm}, we show the radial distribution of low-mass and high-mass satellites of both the transient and star-forming variety. The left panel shows low-mass ($10^8 <M_* /$ M$_\odot \leq 10^{10}$) galaxies, and the right panel shows high-mass ($10^{10}< M_* /$ M$_\odot \leq 10^{13}$). The mean distribution of each plot is also shown for ease of comparison.

If quenching were driven primarily by environmental factors, we would see a radial difference in the star-forming and transient populations. However, for the high-mass bin, we find that the transient and star-forming radial distributions are nearly identical, suggesting that the quenching mechanisms for these galaxies is unlikely to be primarily group-driven. It is a different story for the low-mass galaxies. For these, the transients have a steeper inner radial profile, and are consistent with having been in the massive halo for a long time, while the star-forming galaxies have a radial profile more consistent with recent infall \citep{oman2013, jaffe2018}. This supports the idea that low-mass galaxies are more susceptible to the environmental effects, while the quenching of high-mass galaxies is driven by internal effects.

We can support the role of environmental effects further by recalling the clear halo mass dependence in the ratio of transients to star-forming galaxies seen in Figure \ref{fig:fsfqr}. With increasing halo mass, the proportion of transient galaxies with respect to star-forming galaxies, across all stellar masses, increases. The opposite is true for transient fractions with respect to quenched galaxies. The halo mass dependence for the fraction of transients with respect to star-forming galaxies is further clear evidence that some of the transients are driven by environmental processes. This halo mass dependence is consistent with the work of \citet{paccagnella2019} who also showed a strong dependence of the post-starburst fraction with halo mass. This consistent result, with a very different galaxy selection, clearly points to some role for halo mass in the appearance of this quenching. \par

However, it is important to note, that while there is a strong halo mass dependence for the appearance of transients, even the lowest halo masses show an increasing fraction of transients with respect to star-forming as a function of decreasing radius (e.g., the purple dashed line in the left panel of Figure \ref{fig:fsfqr}).  This strongly suggests that, while the effectiveness of environment-induced quenching is lower in low-mass halos, it is still a factor. This is an intriguing result, which is also consistent with other previous studies \citep{zabludoff1996, paccagnella2019, vulcani2020}. The environmental effect clearly can not be dependent only on halo mass, but perhaps instead some other physical feature which correlates with halo mass, such as relative velocity and intragroup gas density.

We present further evidence for this suggestion in Figure \ref{fig:t_halomass_delv}, which shows the distribution of satellite-central velocity offset with respect to the group velocity dispersion split by low halo mass ($10^{11} < M_h\leq 10^{13.5}$ M$_\odot$; left panel) and high halo mass ($10^{13.5}< M_h\leq10^{16}$ M$_\odot$; right panel). The mean distribution of each plot is also shown for ease of comparison. We find that, for both the low and high halo mass subsets, the transients appear to have lower velocity offsets than do star-forming galaxies in the same mass halos. A lower velocity offset is expected if these galaxies have been in the group for longer than the typical star-forming galaxy. This is evidence that there is a timescale on the order of the group-crossing time before the galaxy becomes a transient galaxy. That said, we see a similar pattern as in Figure \ref{fig:t_lmhm}, where in each mass bin, the transient and star-forming populations have practically identical distributions. This further implies that, while the environmental effects cannot be neglected, they are not the primary driver of quenching mechanisms. \par

Overall, these results describe a picture of the state of star-formation in various classes of galaxies and the confluence of factors which drive them. We see that galaxies which have recently truncated their star-formation have a strong stellar mass dependence, which occurs in both central and satellite galaxies. However, at low stellar mass  ($M_*\lesssim10^{10}$ M$_\odot$), these recently-quenched galaxies are much more likely to be satellite galaxies. Further, the occurrence of these transients shows a strong halo mass and radial dependence, with more star-forming galaxies being transformed into transients in high mass halos and nearer the group center.  We also find that the rejuvenated galaxy candidates do not have an environmental dependence, so will not preferentially revive these transient satellites. Finally, although we find a clear environmental effect responsible for some transients, we still find a sizeable population (especially at high stellar mass) which occur in both central and satellite galaxies in equal occurrences.

\section{Conclusions}\label{sec:conclusion}
By combining H$\upalpha$ and UV measurements, from SDSS and GALEX respectively, we gain a unique insight into the instantaneous star-formation rate of satellite galaxies in galaxy groups. Using properties of the group of each satellite, from the YGC, we examine the environment of the satellite and investigate its significance on quenching. Our main results can be summarised as follows:
\begin{itemize}
\item We find a population of recently-quenched galaxies (`transients') which lack H$\upalpha$ emission but maintain UV emission indicative of star-formation. This is characteristic of a galaxy which has had its star-formation quenched in the last $\sim$ 100 Myrs. The occurrence of these transients is a strong function of stellar mass, and occur in central and satellites at the same rate in high stellar mass galaxies ($M_* >10^{10}$ M$_\odot$). However, at low stellar masses, these transients are more likely to occur in satellite galaxies than in central galaxies.
\item The occurrence of transient galaxies in the satellite population is a strong function of the total host halo mass and projected group-centric distance. Measured by the ratio of transient galaxies to star-forming galaxies, transients are more likely to occur in high mass halos and closer to the group centre. Both of these features are characteristic of environmentally driven processes.
  \item We examined the population of rejuvenating galaxy candidates which show H$\upalpha$ emission but lack UV mission, suggesting that they could have recently begun their star-formation again after a period of quiescence lasting at least 100 Myrs. This population showed no environmental dependence, which suggests that our `transient' satellite galaxies are not simply a normal part of a on-off star-formation cycle in galaxies. 
  \end{itemize}

  The evidence for an environmental role in the production of transients that we see in low-mass galaxies, as well as the halo mass and radial dependence, is clear. This suggests that some environmental process is acting over short timescales to quench at least some of these galaxies. Given the signatures in the radial and velocity distributions of these galaxies, this process is likely to occur not immediately after a galaxy enters the halo. Given the short timescale of the transient phase, and the long timescale after the entering of the halo, our results agree with the `delayed-then-rapid' quenching scenario of \citet{wetzel2012}. \par
  However, the clear environmental role should not be overstated. The strong stellar mass dependence of the transient galaxies is similar in both satellites and centrals, and indeed the highest occurrence rates of transients occur in high stellar mass galaxies and equally in central and satellites. This suggests that there is a significant additional factor driving the quenching of galaxies besides environment. This, along with a further elucidation of the role of environment, AGN and morphology, will be the subject of future work. \par

\section*{Acknowledgements}
The authors thank the anonymous reviewer for their constructive comments. We also thank Felicia Ziparo for discussions at an early stage of this work. CC acknowledges the support of the School of Physics and Astronomy at the University of Birmingham. SLM acknowledges support from the Science and Technology Facilities Council through grant number ST/N021702/1. \par
Funding for the Sloan Digital Sky Survey IV has been provided by the Alfred P. Sloan Foundation, the U.S. Department of Energy Office of Science, and the Participating Institutions. SDSS-IV acknowledges support and resources from the Center for High-Performance Computing at the University of Utah. The SDSS web site is www.sdss.org.

SDSS-IV is managed by the Astrophysical Research Consortium for the 
Participating Institutions of the SDSS Collaboration including the 
Brazilian Participation Group, the Carnegie Institution for Science, 
Carnegie Mellon University, the Chilean Participation Group, the French Participation Group, Harvard-Smithsonian Center for Astrophysics, 
Instituto de Astrof\'isica de Canarias, The Johns Hopkins University, Kavli Institute for the Physics and Mathematics of the Universe (IPMU) / 
University of Tokyo, the Korean Participation Group, Lawrence Berkeley National Laboratory, 
Leibniz Institut f\"ur Astrophysik Potsdam (AIP),  
Max-Planck-Institut f\"ur Astronomie (MPIA Heidelberg), 
Max-Planck-Institut f\"ur Astrophysik (MPA Garching), 
Max-Planck-Institut f\"ur Extraterrestrische Physik (MPE), 
National Astronomical Observatories of China, New Mexico State University, 
New York University, University of Notre Dame, 
Observat\'ario Nacional / MCTI, The Ohio State University, 
Pennsylvania State University, Shanghai Astronomical Observatory, 
United Kingdom Participation Group,
Universidad Nacional Aut\'onoma de M\'exico, University of Arizona, 
University of Colorado Boulder, University of Oxford, University of Portsmouth, 
University of Utah, University of Virginia, University of Washington, University of Wisconsin, 
Vanderbilt University, and Yale University.\par
This paper is partly based on archival data from the 6th General Release. GALEX was operated for NASA by California Institute of Technology under NASA contract NAS-98034.

\section*{Data Availability}
The data used for this work are available at the links listed above in the article.


\bibliographystyle{mnras}
\bibliography{main} 


\appendix
\section{Dust Attenuation}\label{sec:dust}
As stated in \textsection \ref{subsec:quenched}, we do not apply dust attenuation corrections to the star-formation rates. This allows us to use the full range of galaxies with well-measured H$\upalpha$ fluxes, while mainting the connection directly to the massive stars that produce this emission. Given the dust is expected to principally come from local scales, it may be that such effects are similar in centrals and satellites for similar star-formation rates. However, we must test this assumption, and show its consistency in all the regions we probe. We estimate the effect these dust corrections would have on our SFRs by using the Balmer decrement, as in \citet{gilbank2007} (see references therein):
\begin{equation}
    A_{\text{H}\upalpha} = \frac{2.5}{k_{\text{H}\upbeta}/k_{\text{H}\upalpha} -1}\log\Bigg(\frac{1}{2.85}\frac{f_{\text{H}\upalpha}}{f_{\text{H}\upbeta}}\Bigg),
\end{equation}
where $k_{\text{H}\upbeta}/k_{\text{H}\upalpha} =1.48$ and $f_{\text{H}\upalpha(\upbeta)}$ is the H$\upalpha$($\upbeta$) flux. Using the fluxes of H$\upbeta$ and H$\upalpha$ for galaxies with both well measured we find that the average attenuation is $A_{\text{H}\upalpha}\sim 1$, and this shows no difference between satellites and central galaxies. Since applying this attenuation correction would have no appreciable difference on the differential effects of centrals and satellites and thus no effect on our subsequent analysis, we do not apply it. However, we fully investigate the relative effect of dust attenuation here. We use the ratio of H$\upalpha$ to H$\upbeta$ as a proxy for the attenuation coefficient. We ensure that each galaxy that is considered to have well-measured H$\upalpha$ or H$\upbeta$ has a SNR > 3 for both measurements.\par
\begin{figure*}
    \centering
    \includegraphics[width=\textwidth]{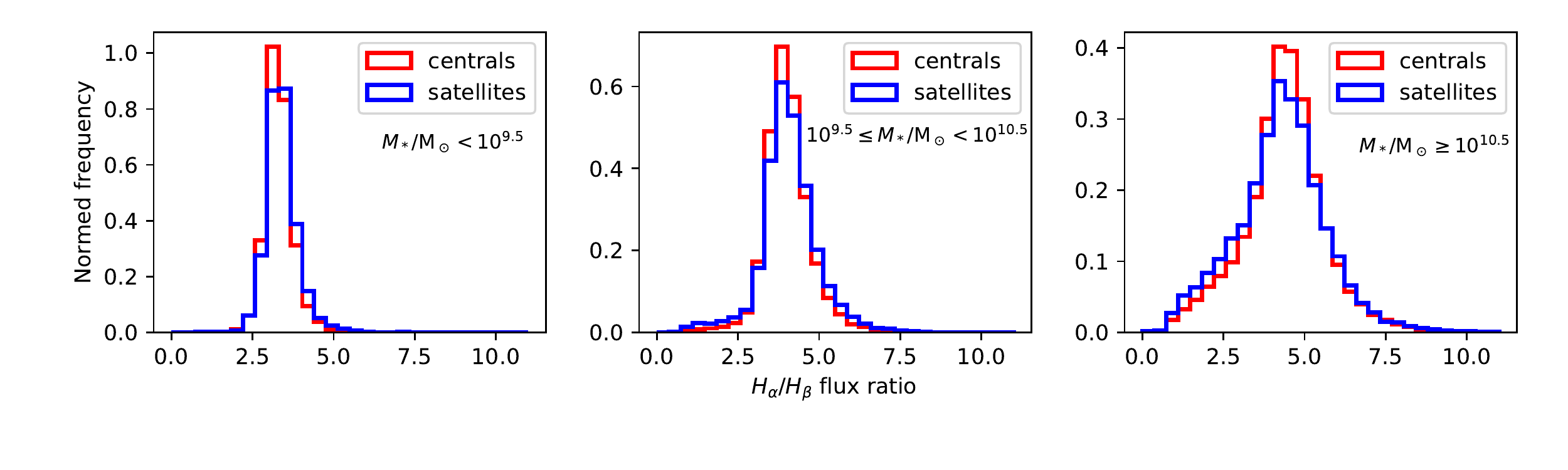}
    \caption{Distributions of flux ratios H$\upalpha/$H$\upbeta$ in three stellar mass ($M_*$) bins as labeled in each panel. We see the flux ratio distribution is the same for central galaxies and satellite galaxies with respect to stellar mass.}
    \label{fig:ratio_m}
\end{figure*}

Figure \ref{fig:ratio_m} illustrates the flux ratio for centrals and satellites in three stellar mass bins, $M_*/$M$_\odot<10^{10}$, $10^{10}\leq M_*/$M$_\odot<10^{11}$ and $M_*/$M$_\odot \geq 10^{11}$. Each bin contains 11,006 centrals and 4,031 satellites, 49,119 centrals and 13,300 satellites, and 63,854 centrals and 11,474 satellites respectively. The mean flux ratio in each bin is $3.32\pm0.03$ (centrals) and $3.40\pm0.05$ (satellites), $4.04\pm0.04$ (centrals) and $4.08\pm0.05$ (satellites), and $4.38\pm0.31$ (centrals) and $4.27\pm0.15$ (satellites) respectively. Apart from negligible differences, the distributions of the flux ratio are identical. This implies that the dust attenuation is the same for centrals and satellites with respect to stellar mass, at least for the case when both H$\upalpha$ and H$\upbeta$ are well measured.\par
\begin{figure}
    \centering
    \includegraphics[width=\columnwidth]{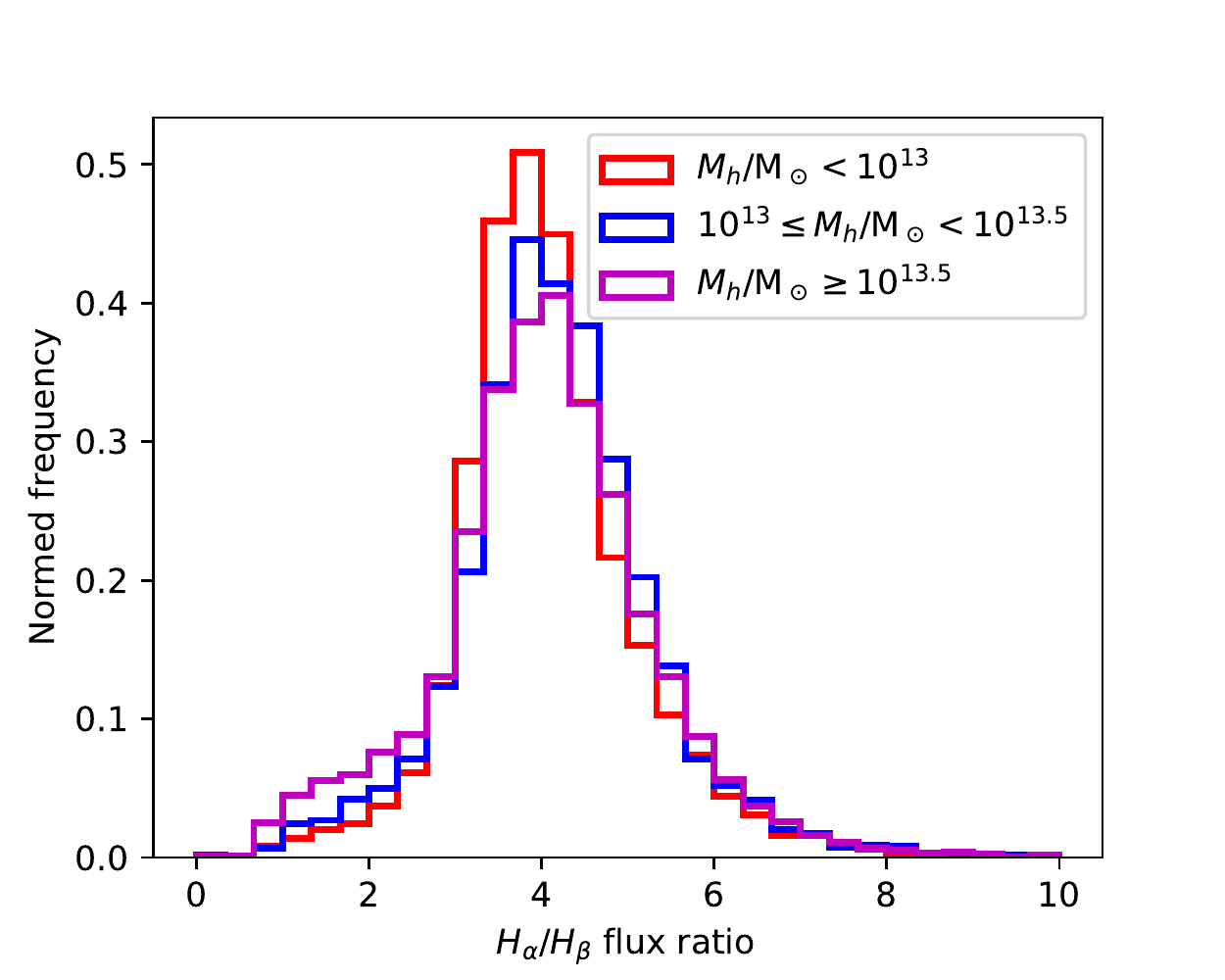}
    \caption{Distributions of flux ratios H$\upalpha/$H$\upbeta$ in three halomass ($M_h$) bins. We see the flux ratio distribution is the same regardless of halomass.}
    \label{fig:ratio_hm}
\end{figure}

To check whether there could be differences in dust attenuations, we will look at subgroups of satellites in different host halo masses, different radial positions and different velocity offsets. First, in Figure \ref{fig:ratio_hm}, we plot the flux ratio with respect to three group halomass bins. These bins have total galaxy numbers of 9,712, 5,483, and 11,266 respectively and the mean flux ratio in each bin is $4.08\pm0.07$, $4.19\pm0.09$, and $4.06\pm0.15$ respectively. These differences are minimal. \par

\begin{figure}
    \centering
    \includegraphics[width=\columnwidth]{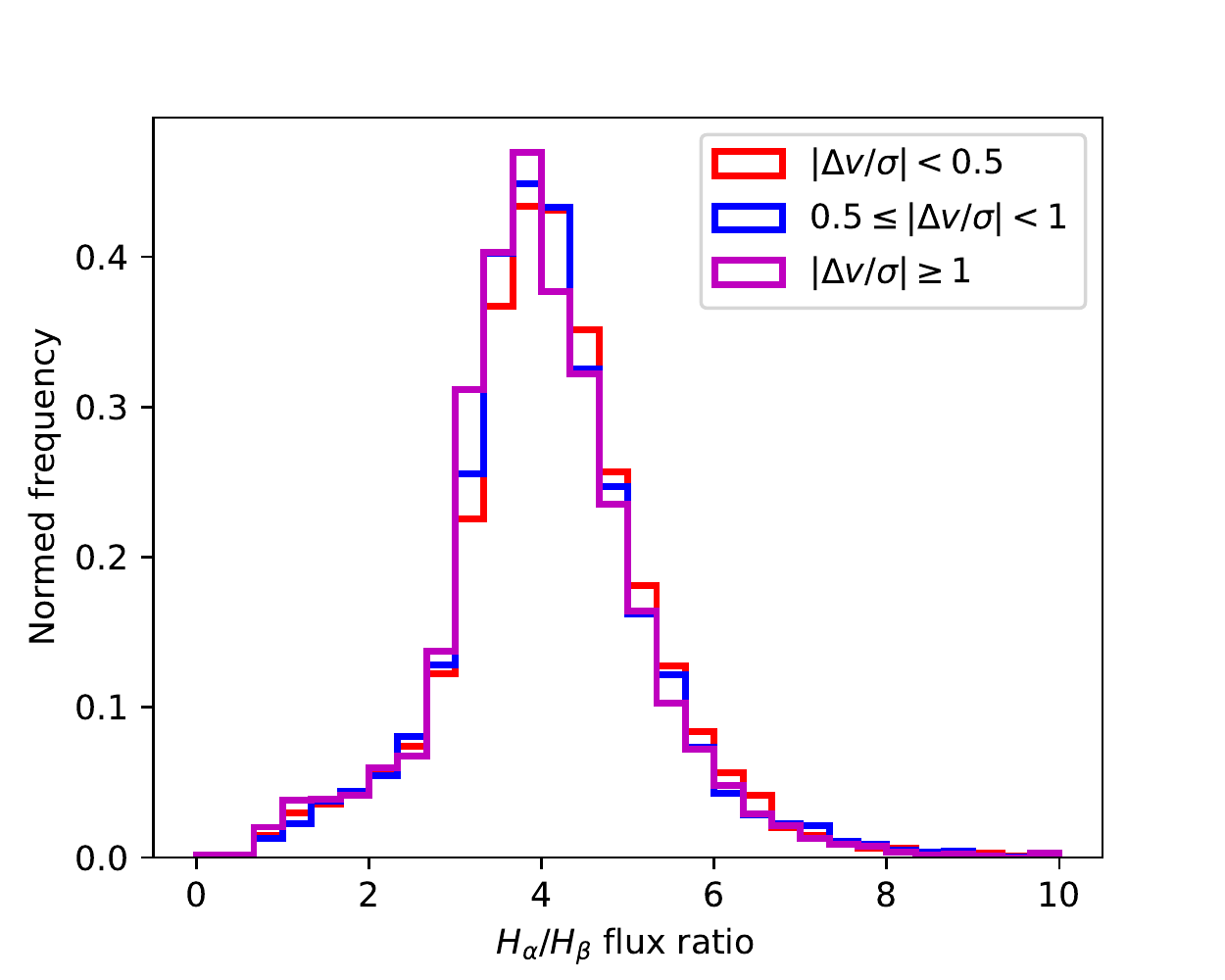}
    \caption{Distributions of flux ratios H$\upalpha/$H$\upbeta$ in three absolute normalised velocity offset ($|\Delta v/\upsigma|$) bins. We see the flux ratio distribution is the same regardless of velocity offset.}
    \label{fig:ratio_v}
\end{figure}

Next, we  plot the flux ratio of satellite galaxies in different bins of absolute normalised velocity offset, $|\Delta v/\upsigma|$. This is shown in Figure \ref{fig:ratio_v}. The number of galaxies in each bin is 14,643, 7,665, and 4,153, with mean flux ratio $4.13\pm0.06$, $4.08\pm0.18$, and $4.01\pm0.09$ respectively.  Again, we see negligible difference in flux ratio.\par

\begin{figure}
    \centering
    \includegraphics[width=\columnwidth]{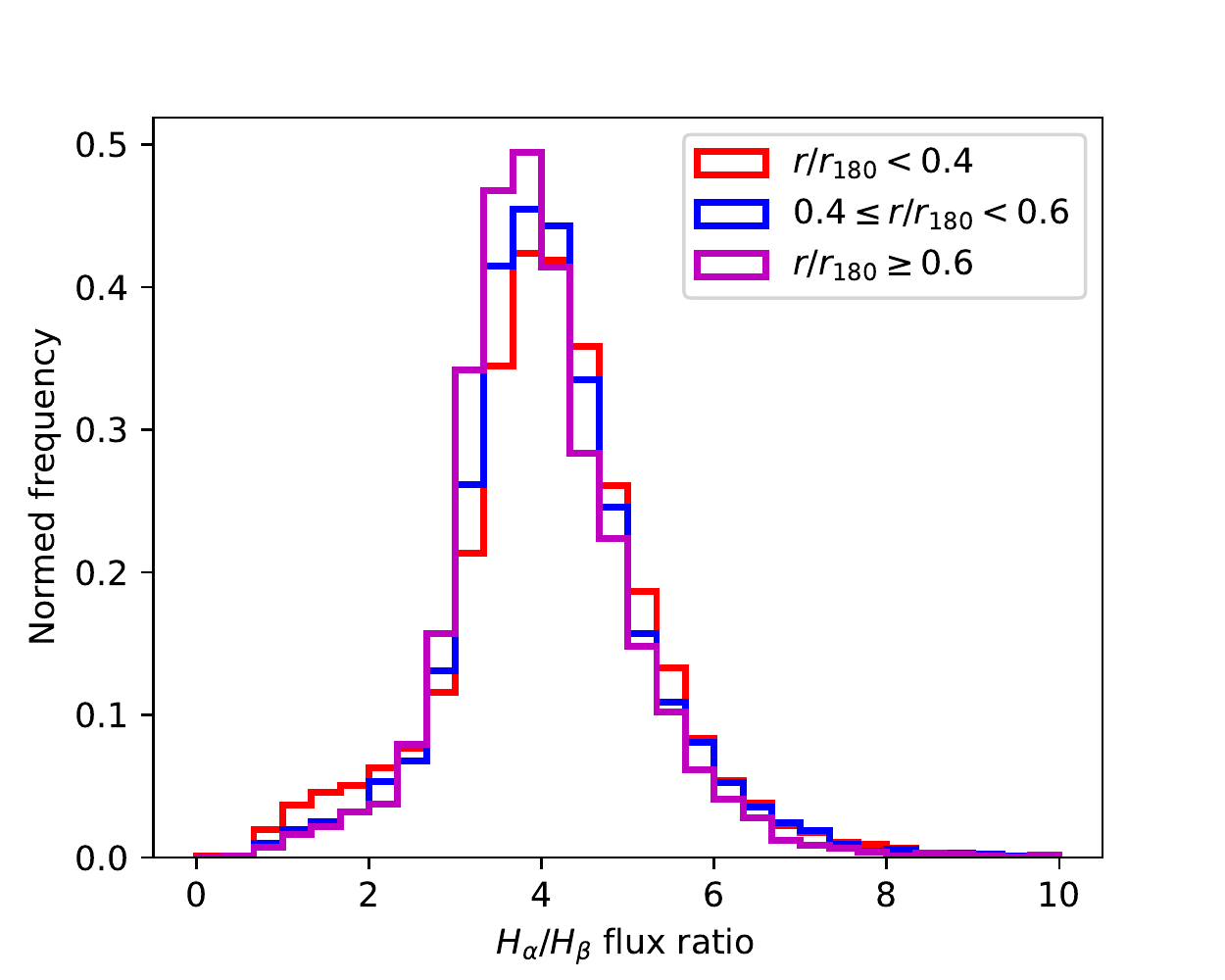}
    \caption{Distributions of flux ratios H$\upalpha/$H$\upbeta$ in three group-centric radius ($r/r_{180}$) bins. We see the flux ratio distribution is the same regardless of group-centric radius.}
    \label{fig:ratio_r}
\end{figure}

Finally, Figure \ref{fig:ratio_r} plots the flux ratio of satellite galaxies in bins of group-centric radius, $r/r_{180}$. The number of galaxies in each bin are 15,811, 5,690, and 4,960, with mean flux ratio $4.12\pm0.13$, $4.11\pm0.09$, and $3.99\pm0.07$ respectively. There is a slightly larger flux ratio (i.e. dust attenuation coefficient) for galaxies closer to the group centre, and a slightly lower flux ratio for galaxies towards the group edge. However, these slight offsets are considered negligible.\par
\begin{figure}
    \centering
    \includegraphics[width=\columnwidth]{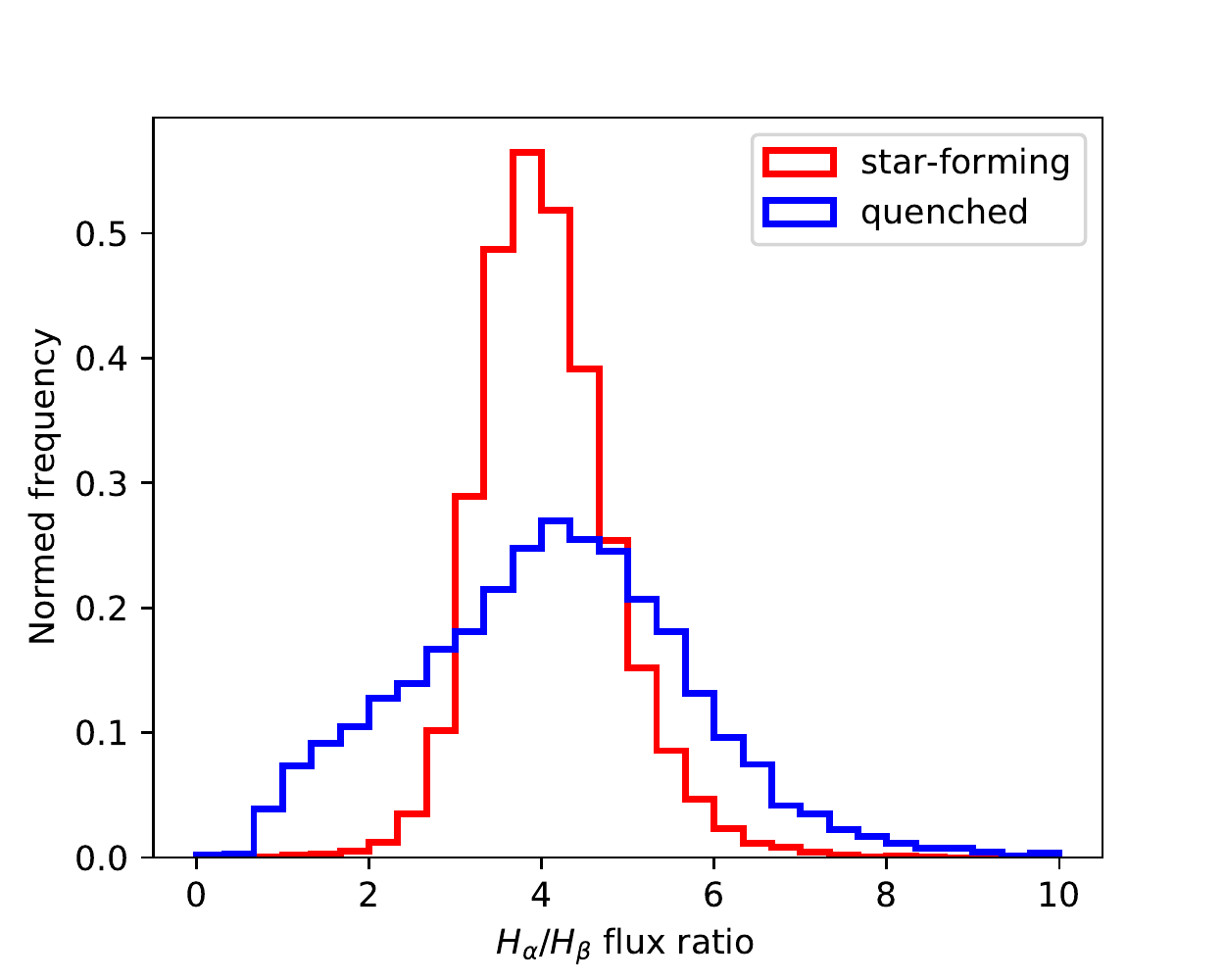}
    \caption{Distributions of flux ratios H$\upalpha/$H$\upbeta$ for star-forming ($\log(\text{sSFR}_{\text{UV}}[\text{yr}^{-1}])>-11.1$) and quenched ($\log(\text{sSFR}_{\text{UV}}[\text{yr}^{-1}])\leq-11.1$) galaxies. We see the quenched population is not unduly biased by high flux ratios (i.e. dust attenuation coefficients).}
    \label{fig:ratio_s}
\end{figure}

It is possible that galaxies that are deemed passive by sSFR have preferentially higher dust attenuation coefficients, meaning their sSFR is underestimated. However, in Figure \ref{fig:ratio_s} we see this is not the case. This figure plots the flux ratios of star-forming and quenched galaxies as these were defined in \textsection \ref{subsec:quenched}. We see that, for quenched galaxies (i.e. low sSFR), there is a wide range of flux ratio available. In fact, the mean flux ratio for quenched galaxies is lower than that of star-forming galaxies, with $3.90\pm0.09$ and $4.16\pm0.11$ respectively. These results confirm that our results are not biased by dust attenuation. \par

In this Appendix, we have shown that there appears to be no evidence for differential dust effects which could be affecting the bulk of our results. However, we caution that we can not rule out effects when H$\upbeta$ can not be accurately measured, or for individual galaxies.

\bsp	
\label{lastpage}
\end{document}